\documentclass[twocolumn,dvipsnames, usenames]{aastex631}
\usepackage[utf8]{inputenc}
\usepackage{amssymb,amsmath,verbatim,mathtools,needspace,enumitem,etoolbox,graphicx,physics,microtype,ulem,breqn}
\usepackage{xspace} 
\usepackage{booktabs}
\usepackage{upgreek} 
\usepackage{acronym}
\usepackage{url}

\newcommand{\MKI}{\affiliation{Department of Physics and Kavli Institute for Astrophysics and Space Research, Massachusetts Institute of Technology, 77 Massachusetts Ave, Cambridge, MA 02139, USA}}

\defcitealias{tic_184}{Z24}
\defcitealias{reed_tipped_2005}{R05}

\date{\today}
\submitjournal{ApJ}

\begin{document}

\title{TIC 435850195: The Second Tri-Axial, Tidally Tilted Pulsator}

\correspondingauthor{Rahul Jayaraman}
 \email{rjayaram@mit.edu}
\author[0000-0002-7778-3117]{Rahul Jayaraman} \MKI
\author[0000-0003-3182-5569]{Saul A. Rappaport}\MKI
\author[0000-0003-0501-2636]{Brian Powell}\affiliation{NASA Goddard Space Flight Center, 8800 Greenbelt Road, Greenbelt, MD 20771, USA}
\author[0000-0001-7756-1568]{Gerald Handler}\affiliation{Nicolaus Copernicus Astronomical Center, Polish Academy of Sciences, ul. Bartycka 18, PL-00-716 Warszawa, Poland}
\author{Mark Omohundro}\affiliation{Citizen Scientist, c/o Zooniverse, Dept., of Physics, University of Oxford, Denys Wilkinson Building, Keble Road, Oxford OX1 3RH, UK}
\author{Robert Gagliano}\affiliation{Amateur Astronomer, Glendale, AZ 85308, USA}
\author[0000-0001-9786-1031]{Veselin Kostov}\affiliation{NASA Goddard Space Flight Center, 8800 Greenbelt Road, Greenbelt, MD 20771, USA}\affiliation{SETI Institute, 189 Bernardo Avenue, Suite 200, Mountain View, CA 94043, USA}
\author[0000-0002-4544-0750]{Jim Fuller}\affiliation{TAPIR, Mailcode 350-17, California Institute of Technology, Pasadena, CA 91125, USA}
\author[0000-0002-1015-3268]{Donald W. Kurtz}\affiliation{Centre for Space Research, North-West University, Mahikeng 2745, South Africa}\affiliation{Jeremiah Horrocks Institute, University of Central Lancashire, Preston PR1 2HE, UK}
\author{Valencia Zhang}\affiliation{Phillips Academy, Andover, MA 01810, USA}
\author[0000-0003-2058-6662]{George Ricker}\MKI

\begin{abstract}
The Transiting Exoplanet Survey Satellite (TESS) has enabled the discovery of 
numerous tidally tilted pulsators (TTPs), which are pulsating stars in close binaries 
where the presence of a tidal bulge has the effect of tilting the primary star's 
pulsation axes into the orbital plane. Recently, the modeling framework developed to 
analyze TTPs has been applied to the emerging class of tri-axial pulsators, which 
exhibit nonradial pulsations about three perpendicular axes. In this work, we report on the
identification of the second-ever discovered tri-axial pulsator, with 
sixteen robustly-detected pulsation multiplets, of which fourteen are dipole 
doublets separated by 2$\nu_{\rm orb}$. We jointly fit the spectral energy distribution 
(SED) and TESS light curve of the star, and find that the primary is slightly evolved
off the zero-age main sequence, while the less massive secondary still lies 
on the zero-age main sequence. Of the fourteen doublets, we associate eight with 
$Y_{10x}$ modes and six with novel $Y_{10y}$ modes.  We exclude the existence 
of $Y_{11x}$ modes in this star and show that the observed pulsation modes must 
be $Y_{10y}$. We also present a toy model for the tri-axial pulsation framework 
in the context of this star. The techniques presented here can be 
utilized to rapidly analyze and confirm future tri-axial pulsator candidates.
\end{abstract}



\section{Introduction}

The Transiting Exoplanet Survey Satellite (TESS; \citealt{ricker_tess})
has revealed diverse classes of pulsating stars, including the novel class of ``tidally 
tilted pulsators'' (TTPs), wherein pulsating stars in tight binaries (with periods typically 
less than $\sim$\,2\,d) have their pulsation axes tilted into the orbital plane 
by the tidal bulge induced by their companions. Specifically, in these systems,
the pulsation axis is aligned with the tidal bulge, rather than with the spin axis of the star. 
To date, there have been a handful of tidally tilted pulsators discovered
\citep{hd_74420,co_cam,tic_6332,hd_265435,tic_184,jennings_kic_485_ttp}, 
and several other candidates reported in the literature
(see, e.g., \citealt{candidate_ttp}).

A key property of tidally tilted systems is that the observer is able to study the star 
through a wide range of latitudinal angles (from 0$^\circ$ to 360$^\circ$) with respect 
to the pulsation axis. {Tidally tilted modes exhibit amplitude and phase modulations 
around the orbit, thereby leading to splitting of the modes, seen in periodograms
of their light curves.} Unlike typical pulsators, where the observer's viewing angle 
remains constant with respect to the pulsation axis, tidally tilted pulsations
yield a unique perspective on pulsating stars as they orbit their respective companions.

{``Tidally perturbed” pulsators, which also exhibit oscillation 
modes whose amplitudes are modulated over the course of the orbit, 
could also be interpreted as less extreme manifestations of the tidal 
tilting phenomenon (see, e.g., \citealt{v453_cyg_perturbed,v456_cyg,
g_mode_perturbed,u_gru_perturbed,kic_985_perturbed}). However, there 
may also be other factors at play in these stars causing the observed 
changes in the pulsation modes, such as asynchronous rotation 
(see, e.g., the Introduction of \citealt{u_gru_initial} and references 
therein) or a tidal amplification mechanism, described in \citet{ttp_modeling}.}

Significant modeling of the first three tidally tilted pulsators that were discovered was 
performed by \citet{ttp_modeling}; a unique attribute of the
tidal tilting phenomenon is that the pulsation modes are able to be identified with specific
$\ell$ and $|m|$ values, providing valuable insights into both the star's 
evolutionary state and its interior structure (as was shown in \citealt{hd_265435}). Initially, the
framework for understanding TTPs was based upon the oblique pulsator model, 
developed by \citet{roap_oblique_pulsator} for the rapidly oscillating Ap (roAp) stars;
in this scheme, the gravitational field of the companion star in TTPs plays an 
analogous role to the magnetic field intrinsic to roAp stars. However, it remains an 
open question why only certain tight binaries experience tidal tilting
or, for that matter, why certain modes in a given star can be 
tidally tilted, while others in the same star are not.

Recently, \citet{tic_184} (hereafter \citetalias{tic_184}) reported the 
unique discovery of TIC\,184743498, the first 
``tri-axial pulsator'' (TAP). This star was found to pulsate along three different
axes, and Section 8 of \citetalias{tic_184} presented a perturbative model for how 
modes coupled by the tidal bulge could naturally produce pulsations along three 
different axes. This work presents the 
second TAP discovered with TESS. Section \ref{sec:obs} presents the
details of the TESS data used in the analysis, and our initial processing of the light
curve. Section \ref{sec:eb} presents our joint fit to
the system's light curve and spectral energy distribution (SED), and enumerates the 
system parameters resulting from that analysis. Section \ref{sec:puls} presents 
an asteroseismic analysis of 
the system, and Section \ref{sec:disc} presents a toy model of the tidal tilting
framework for this system, while exploring (and refuting)
alternative explanations for this star's behavior.

\section{Observations and Data Processing}
\label{sec:obs}

\begin{figure} 
    \includegraphics[width=\linewidth]{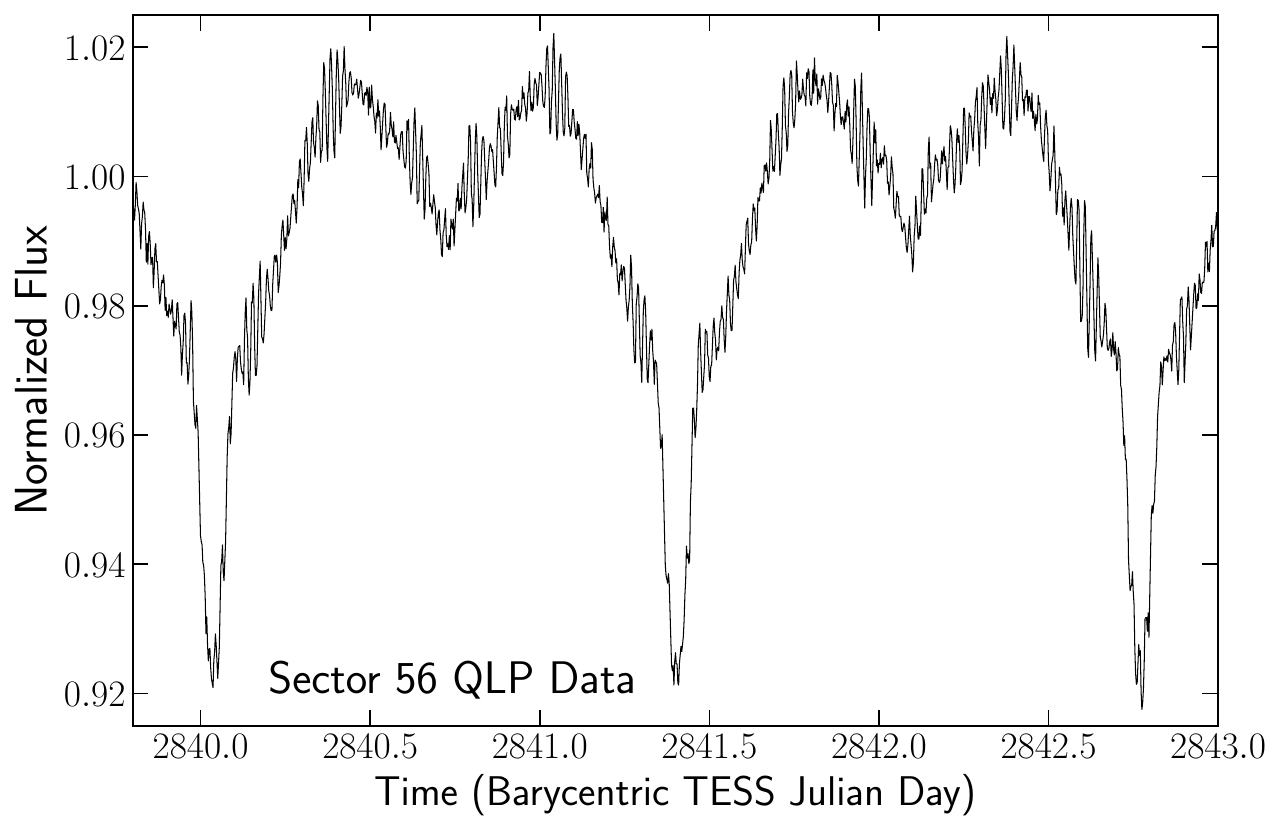}
    \caption{A $\sim$3.5\,d portion of the TESS light curve of TIC 435850195
    from the Quick-Look Pipeline \citep{qlp_i, qlp_ii}, sampled at 200\,s 
    cadence. The amplitudes and phases of most of the pulsation modes are 
    seen to vary systematically with the orbital phase. There are also prominent
    ellipsoidal light variations (discussed further in Section 
    \ref{subsubsec:elv_var}) and prominent primary eclipses.}
\end{figure}

\begin{figure}
    \centering
    \includegraphics[width=\linewidth]{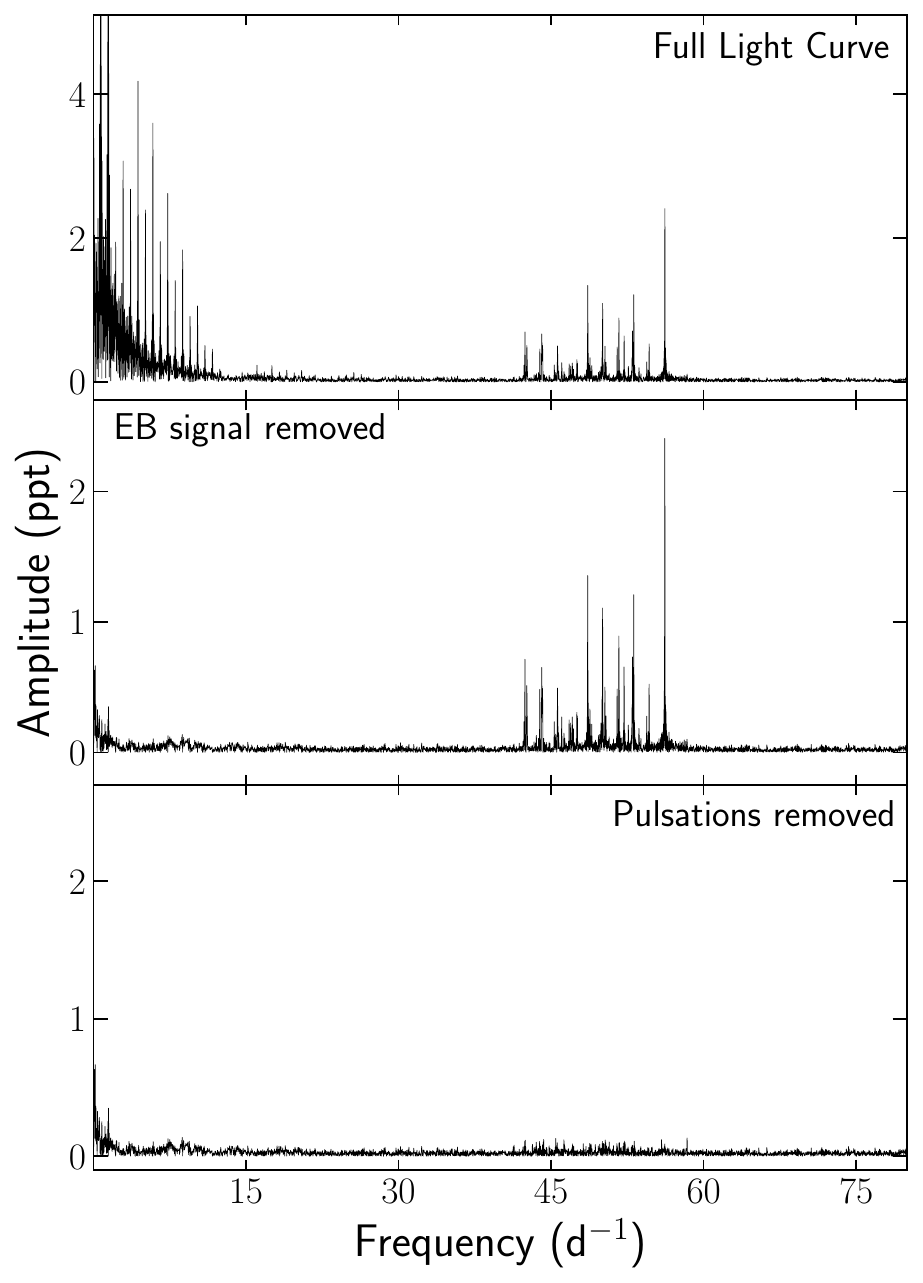}
    \caption{The Discrete Fourier Transform (DFT; \citealt{kurtz_dft}) of
    TIC 435850195, with sequential subtractions of the eclipsing binary
    signal in the raw light curve (visible in the top panel) and the
    $\delta$\,Scuti pulsations (visible in the middle panel). Each set
    of periodic signals was fit for using least-squares and then subtracted
    from the light curve; the bottom panel shows the DFT of the residual 
    light curve, after both the EB signal and pulsations were fitted and removed.
    Note the differing scales between the top panel and the lower two.}
    \label{fig:stack_ft}
\end{figure}

A team of citizen scientists, including MO and RG, have been conducting 
a visual survey of light curves from the TESS full-frame images (FFIs); more 
information about this effort can be found in \citet{kristiansen_vsg}. 
One major goal of this survey is to find unique stars in eclipsing binary (EB) systems. 
To this end, \citet{powell_2022_ffi_lcs} generated a set of millions of EB light curves 
from the TESS full-frame images using the {\tt eleanor} pipeline \citep{eleanor}
and a machine learning approach, detailed in Section 2 of \citet{powell_2021}.
TIC\,435850195 was initially identified as part of this 
effort, during a review of EB light curves identified by the 
neural network from TESS Sector\,56. This star was 
observed for the first time in Sector\,56 at 200\,s cadence.
Further details about the system are given in Table \ref{tab:system_params}.

Follow-up analyses utilized the 200\,s-cadence light curve from the MIT 
Quick-Look Pipeline (QLP; \citealt{qlp_i, qlp_ii}). The QLP light curve 
was downloaded using {\tt lightkurve} \citep{lightkurve}, and we used the 
orbital period calculated by the Gaia mission
\citep{gaia_mission,gaia_dr3}, reported in Table 
\ref{tab:system_params} (Siopis et al., in prep), to analyze the 
eclipses. Orbital period estimates from both the TESS data and 
archival data from the All-Sky Automated Survey for Supernovae 
\citep{asassn} were consistent with the Gaia value. Using the TESS 
data, we also calculated a value for the ephemeris $t_0$, using the
time of the first primary eclipse. We find that 
$t_0$ = BJD\,2459826.36448 (where BJD is
``Barycentric Julian Date'' in the TDB system, as defined in 
\citealt{eastman_bjd}).

After downloading the normalized light curve, we fit for 50 harmonics of 
the orbital period, along with a constant offset, to the light curve. This procedure 
was used to reconstruct the eclipsing  binary light curve 
for subsequent analysis with our light curve fitting code.
We used the residuals from this fit as the ``pulsational'' light curve
for our asteroseismic analysis in Section \ref{sec:puls}.  
With this ``pure'' pulsational light curve, we then fit for the most significant
frequencies, including their amplitudes and phases. High-amplitude singlet
frequencies, and those frequencies that formed part of a multiplet, are
enumerated, along with their best-fit amplitude and phase, in Table \ref{tab:freqs}.
The Discrete Fourier Transforms \citealt{kurtz_dft}
of each light curve (raw QLP light curve, ``pure'' pulsational 
light curve, and the residuals after all the frequencies are subtracted),  
are shown in Figure \ref{fig:stack_ft}.

The spectral energy distribution was downloaded from the Vizier SED 
viewer\footnote{\url{http://vizier.cds.unistra.fr/vizier/sed/}} \citep{vizier}. 
We used the photometric measurements of this system available from the
following sources:
Gaia \citep{gaia_mission,gaia_dr3}, Pan-STARRS1 \citep{ps1_survey}, 
the Wide-field Infrared Survey Explorer \citep{wise}, the 2-Micron
All-Sky Survey \citep{2mass_catalog,2mass_survey}, the Tycho-2 catalogue
from the Hipparcos survey \citep{tycho2_catalog}, the Sloan Digital Sky
Survey \citep{sdss_dr16}, and the Galaxy Evolution Explorer ultraviolet
source catalog \citep{galex_catalog}.

\begin{table}
    \centering
    \begin{tabular}{l l c}
        \hline 
        \hline
       Parameter  & Value & Reference  \\
       \hline
        Right Ascension (h m s) & 22:54:51.96 & (1)\\
        Declination ($^\circ$\,'\,'') & 20:47:52.17 & (1) \\
        $T_{\rm mag}$ & 10.551\,$\pm$\,0.009& (2)\\
        $G_{\rm mag}$ & 10.7251\,$\pm$\,0.0009 & (1)\\
        $G_{\rm BP} - G_{\rm RP}$ & 0.3493\,$\pm$\,0.0043 & (1)\\
        $K_1$ (km\,s$^{-1}$) & 46.13\,$\pm$\,2.61& (3) \\
        Distance (pc) & 521\,$\pm$\,10& (1) \\
        Parallax (mas) & 1.801\,$\pm$\,0.034& (1) \\
        $\mu_{\rm RA}$ (mas\,yr$^{-1}$) & $-16.91$\,$\pm$\,0.03 & (1)\\
        $\mu_{\rm Dec}$ (mas\,yr${-1}$) & $-12.97$\,$\pm$\,0.04& (1) \\
        Period (d) & 1.36719\,$\pm$\,0.00006 & (1, 4) \\
        $t_0$ (BJD--2\,457\,000) & 2826.36448 & This work \\ 
         \hline 
    \end{tabular}
    \caption{Information about the TIC 435850195 system.
    The reference numbers in the third column correspond to:
    (1) \citet{gaia_mission,gaia_dr3}; (2) \citet{tic_stassun};
    (3) \citet{gaia_rvs}; (4) Siopis et al. (in prep).}
    \label{tab:system_params}
\end{table}

\section{Estimating Binary System Parameters}
\label{sec:eb}

We simultaneously fit the orbital light curve and the composite SED of TIC 435850195
using a custom Markov Chain Monte Carlo (MCMC) fitting code.  There are 14 (or 16) free 
parameters in the fit: the two masses ($M_1$, $M_2$), the system age, the orbital 
inclination angle $i$, 8 physically 
motivated Fourier components of the light curve (and an additional
constant offset), the distance to the system, and the line-of-sight extinction $A_G$.
The aforementioned eight Fourier components correspond to the first four orbital harmonics, and
represent a combination of the ellipsoidal light variations (ELVs), illumination 
effects, and any co-rotating starspots; physical explanations for these components
are given in Table \ref{tab:elv_params}.

To characterize this system, we make use of the MIST stellar evolution 
tracks \citep{mist_i,mist_ii}, which yield
the stellar radii ($R_1$, $R_2$), effective temperatures ($T_{\rm eff, 1}$; 
$T_{\rm eff, 2}$), and luminosities ($L_1$, $L_2$) as a function of the stellar masses and
system age. As a result, at each step of the MCMC, we are able to utilize the masses
to calculate the orbital separation $a$ (from the period and Kepler's
third law), allowing us to compute the eclipse geometry. We can also use
the stellar radii and effective temperatures to model the composite SED, using the
model atmospheres of \citet{castelli_kurucz}. As part of our analysis, we assume that the two stars have evolved in a coeval manner
since their formation---in particular, we assume that there have been no prior episodes
of mass transfer between them.

The final possible free parameters are the distance to the source and the line-of-sight
extinction $A_G$. We first ran an MCMC with both the distance and the extinction as free 
parameters, to ensure agreement with the best-fit values from Gaia. Then, 
we fixed the distance to 
521\,pc, and the extinction to 0.42, and re-ran the fit. For each value in the SED, 
we corrected for extinction using the \citet{cardelli_extinction} 
extinction law and this $A_G$ value.

In the following subsections, we describe in more detail the simplified orbital 
light curve model and the SED fit. The MCMC was run with {ten parallel walkers, for 
175\,000 steps. We discarded the first 20\,000 samples in each chain as burn-in.}


\subsection{Light Curve Fitting}

We performed a fit to the folded and binned Fourier-reconstructed light curve. 
This light curve was generated using the coefficients from the fit to 50 orbital 
harmonics described in Section \ref{sec:obs}, and was binned to 2\,min. This
approach was taken in order to minimize any leakage of signals
from the pulsation frequencies into the ``pure'' eclipsing binary 
light curve. Our simplified light curve fitting code incorporates three parts:
(i) eclipses from two spherical stars\footnote{The primary star in this system
fills 50\% of its Roche lobe, and the secondary $\sim$\,30\% (for details, see
Section \ref{subsec:sys_params}). Such stars, in the Roche potential, are found
to be spherical to within $\lesssim$2\% in all dimensions, and to within $\lesssim$1\%
in the dimensions relevant during eclipses---i.e., the $y$ and $z$ axes
perpendicular to the tidal axis.}; (ii) limb darkening, for which 
we utilize the approximation scheme described in Appendix \ref{app:ld}; and
(iii) a series of four sine and four cosine terms to capture out-of-eclipse
variability, including ellipsoidal light variations, 
the illumination effect, and starspots that are co-rotating with the
orbit. The bolometric correction for the TESS band, which is a function of the
stellar effective temperature,
was approximated by assuming a blackbody spectrum and then 
calculating the ratio of the integrated flux between 6\,000\,\AA\,and 
10\,000\,\AA\,to the total integrated flux across all wavelengths.
Bolometric correction values were calculated for blackbodies having
temperatures between 4\,000\,K to 10\,000\,K.

\begin{figure} 
    \centering
    \includegraphics[width=\linewidth]{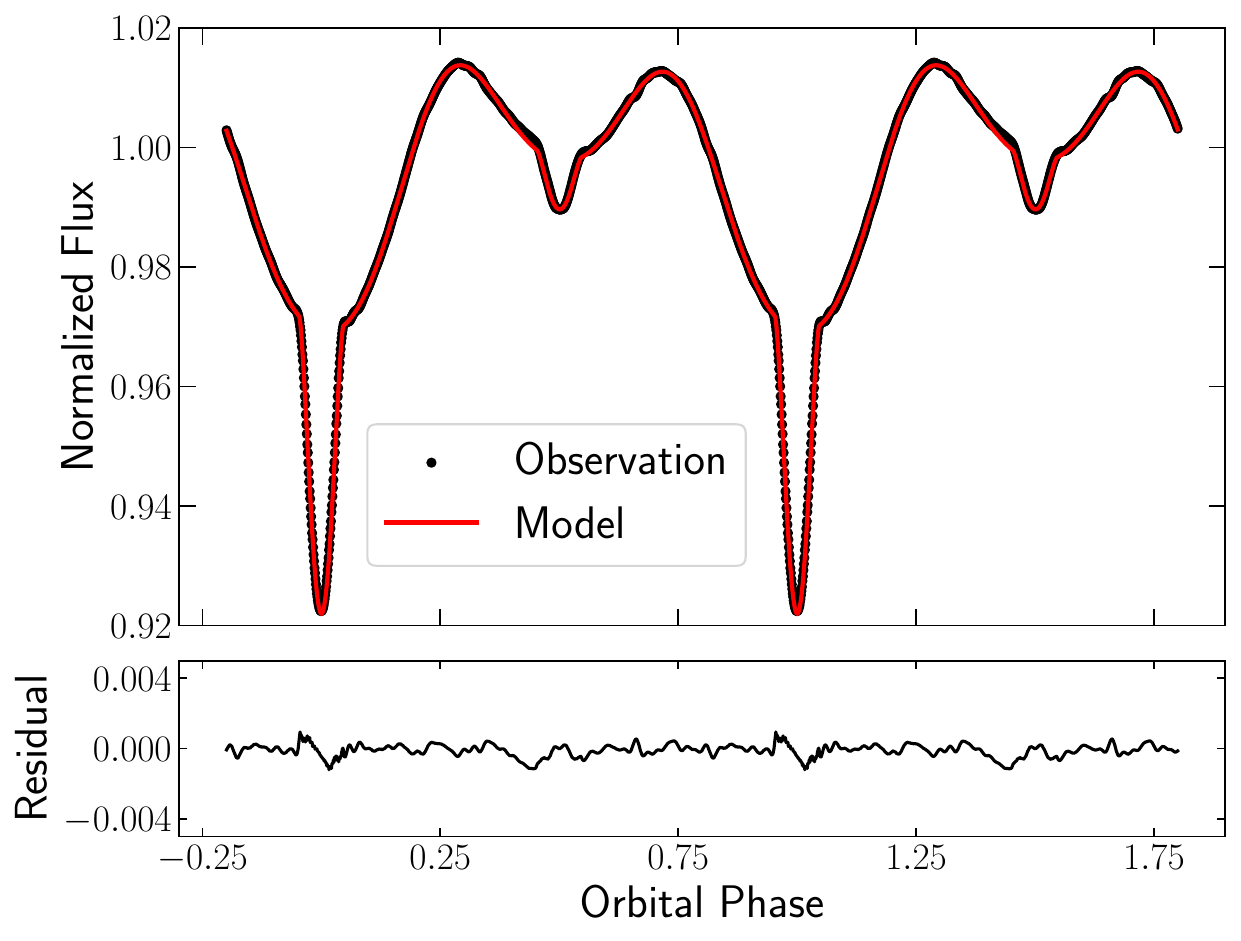}
    \caption{Our best-fit model to the binned, folded, 
    Fourier-reconstructed light curve of TIC 435850195, 
    along with residuals. The 
    residual structure around the times of eclipse 
    could be partially explained by our use of fixed
    limb-darkening coefficients, rather than letting them be
    free parameters.}
    \label{fig:lc-mod-resid}
\end{figure}

\subsubsection{Eclipse Modeling}

At each link in the MCMC chain, we use the stellar radii and
luminosities for the eclipse modeling. We set the
baseline flux as the sum of the two luminosities, multiplied by
their respective bolometric corrections. We then calculated the 
sky-projected separation between the two stars' centers for the current 
values of the orbital inclination angle $i$ and semi-major axis $a$, which
then allowed us to determine the overlapping area between 
the two stars as a function of orbital phase. 

During the portions of overlap, we utilized a one-dimensional integral 
to calculate the limb darkening (LD) flux variations for the 
primary star. We used a quadratic LD model and the coefficients 
from \citet{claret_ld}. We numerically integrated the limb-darkening 
over the overlapping area, in a scheme that is a simplification of
the standard approach from \citet{mandel_agol}. The
relatively simple geometry of this system allows for
a large number of the cases (III-XI) in \citeauthor{mandel_agol} 
to be omitted for the case of the primary eclipse.
Further details of our approach
to calculating the limb darkening can be found in Appendix 
\ref{app:ld}. For the secondary star, whose
limb darkening has a significantly smaller effect on the light curve
shape, we calculated an ``average'' limb-darkening coefficient
for the overlap region and then applied this across that 
entire area; this approach provided a reasonable fit
(see residuals in Figure \ref{fig:lc-mod-resid}).

For the primary, we used the limb-darkening coefficients for a star 
with $T_{\rm eff}=7500$\,K, $\log\,g=3.5$, and microturbulent 
velocity $\xi$\,=\,2\,km\,s$^{-1}$. For the
secondary, we used the coefficients for a star with $T_{\rm eff}=4250$\,K,
$\log\,g=4.0$, and microturbulent velocity $\xi=0$\,km\,s$^{-1}$.
We recognize that there exists a significant debate in 
the exoplanet community over whether limb-darkening 
coefficients should be fixed or free parameters when fitting
transit light curves
(see, e.g., \citealt{csimadia_transit,espinoza_exoplanet_ld}).
Tests with LD coefficients as free parameters did not 
yield a significant improvement in the residuals, so we fixed
them to the \citeauthor{claret_ld} values.

Two orbital cycles of the Fourier-reconstructed light curve, with 
our best-fitting model overlaid (including the out-of-eclipse
fit described in Section \ref{subsubsec:elv_var}), are shown in Figure 
\ref{fig:lc-mod-resid}. Our model fits the light curve
rather well, with a maximum residual of approximately 1 ppt. 

\subsubsection{Out-of-Eclipse Variability}
\label{subsubsec:elv_var}
\begin{table}
    \centering
    \begin{tabular}{l r c}
    \hline 
    \hline
       Coefficient  & Value (ppm) & Physical Explanation  \\
       \hline
        $a_1$ &  $-15178\,\pm\,28$ & Spots + ELV\\
        $a_2$ &  $-13870\,\pm\,41$ & ELV (dominant) \\
        $a_3$ &  $846\,\pm\,19$ & Spots + ELV \\
        $a_4$ &  $1233\,\pm\,23$ & Spots\\
        $b_1$ &  $-83\,\pm\,8$ & Doppler boosting?\\ 
        $b_2$ &  $-1229\,\pm\,10$ & Spots \\ 
        $b_3$ &  $-97\,\pm\,10$ & Spots \\ 
        $b_4$ &  $-15\,\pm\,12$ & Spots \\
        $c_0$ & $-3977\,\pm\,25$ & Offset from 0 \\
        \hline
    \end{tabular}
    \caption{Best-fit amplitudes to our semi-empirical model of the
    out-of-eclipse variability. Two of these parameters ($a_2$ and $b_1$)
    correspond
    to physical attributes of the system, and are further discussed
    in Section \ref{subsubsec:elv_var}. Other parameters 
    account for co-rotating spots on the stellar surface.}
    \label{tab:elv_params}
\end{table}

For the eclipse modeling, we assumed a flat out-of-eclipse
baseline that was normalized to 1. However, in reality, 
light curves of tight binaries exhibit significant out-of-eclipse
variability, for the reasons described earlier. As a result,
we added in a semi-empirical model with 
nine free parameters that has
the functional form, {which is similar to Equation 8 from \citet{carter_wd_elv}}:
\begin{equation}
   c_0 + \sum_{n=1}^{4} a_n\cos{n\omega_{\rm orb} t} + b_n\sin{n\omega_{\rm orb} t}.
\end{equation}
The values of these parameters and their physical significances
are enumerated in Table \ref{tab:elv_params}. With these best-fit 
parameter values, we can attempt to constrain certain physical
attributes of the system. 

\paragraph{Ellipsoidal Light Variations} Stars in tight binaries are
distorted into ellipsoids by the tidal forces from their companion 
({a more detailed discussion can be found in Section IV.2 of
\citealt{kopal_elv}}). This affects the amount of flux that we observe,
due to the fact that both the cross-sectional area from
which the flux appears to be emitted and the gravity darkening
vary throughout the orbit with a frequency of $2\nu_{\rm orb}$.
These contribute to the dominant term in the ELV ($\cos\,2\omega t$);
terms with frequencies
$\nu_{\rm orb}$ and $3\nu_{\rm orb}$ also contribute to the ELV,
but to a much lesser degree.
Equations 14--16 from \citet{carter_wd_elv} relate the amplitude
of each cosine term ($a_1$, $a_2$, and $a_3$ in our notation) 
to physical parameters. 

The ELVs are analytically related to the
physical properties of the system through the linear limb-darkening
coefficient $u$, and the gravity-darkening exponent $y$
\citep{vz_gravdark}. For TIC 435850195, we use the $u$ and $y$ 
values derived for the TESS band 
by \citet{claret_ld}; these are $0.4114$ and 
$0.1245$, respectively. The equation relating $a_2$ and the physical attributes 
of the system is
\begin{equation}
    \label{eq:a2}
    a_2 \simeq -Z_1(2)\,q\left(\frac{R_1}{a}\right)^3\,\sin^2\,i.
\end{equation}
We present below a simplified version of the expression for $Z_1$ given in 
\citet{morris_elv}. Our expression does not include the $k_1$ term in the
original equation that 
accounts for precession induced by a third body.
\begin{equation}
    Z_1(2) = \frac{45 + 3u}{20(3-u)}(1+y) = 1.004 
\end{equation}
The dominant source of uncertainty in this expression arises from
the uncertainty in the coefficients $u$ and $y$. 
We substitute our value of $Z_1(2)$ into Equation \ref{eq:a2}, along
with the relevant best-fit parameters from Table 
\ref{tab:fit_system_params}. Using our calculated value of $a = 6.96$\,R$_\odot$,
we find that $a_2\simeq$\,9813\,ppm. This is $\sim$70\% of
the best-fit value for $a_2$ in Table \ref{tab:elv_params}, suggesting 
that the ELV contribution is the dominant term for the component of the out-of-eclipse
variability with a frequency twice that of the orbit.

\paragraph{Doppler Boosting} Of particular interest is the
$b_1\,\sin\,{\omega t}$ term, which corresponds to the
Doppler boosting effect in tight binaries first detected
by \citet{maxted_db}; a theoretical treatment of this 
effect is given in \citet{loeb_gaudi_doppler}. Doppler
boosting is caused by the motion of the primary inducing
three key effects, all of which are comparable in magnitude:
An increased rate of photon arrivals,
an increase in the net energy of the emitted photons, and
slight relativistic beaming due to the motion of the star.
Using {\tt synphot} \citep{synphot}, we calculated
the Doppler boosting coefficient 
$\alpha_{\rm TESS}$ to be 2.89 for a 7800\,K star (assuming
a blackbody spectrum); for a Vega-type
star, this coefficient increases to 2.92. 

The Doppler boosting amplitude is given by
\begin{equation}
A_{\rm DB} = \alpha \frac{K}{c},
\end{equation}
where $\alpha$ is the coefficient calculated previously. 
If we were to use the Gaia value for $K_1$ (46 km\,s$^{-1}$),
we expect the coefficient $b_1$ of the $\sin \omega_{\rm orb}t$ term 
to be +443 ppm. By comparison, Table \ref{tab:elv_params} shows a value for
$b_1 = -83$ ppm. This discrepancy can be explained by invoking co-rotating
starspots, as these can strongly affect the amplitude of low-frequency
orbital Fourier terms. We can gauge how large these might be by 
looking at terms $a_4$, $b_2$, $b_3$, and $b_4$, where there 
should be minimal to no contribution from physical effects such 
as ellipsoidal light variations, illumination effects, and Doppler 
boosting. At least two of these amplitudes ($a_4$ and $b_2$) are large
enough to interfere with the Doppler boosting term ($b_1$), if they
occur at the orbital frequency. Thus, we conclude that the Doppler
boosting term is undetectable here.  

An interesting question is whether spots in this system can account 
for amplitudes of $\gtrsim 1000$ ppm at the lower orbital 
harmonics. The secondary, which is cool enough to have large 
spots, contributes $\sim$1\% of the system light. To produce 
signals as large as 1000 ppm requires spots of 10\% amplitude 
on the secondary, which is plausible.

\subsection{SED Fitting}
\label{subsec:sed_fitting}

\begin{figure} 
    \centering
    \includegraphics[width=\linewidth]{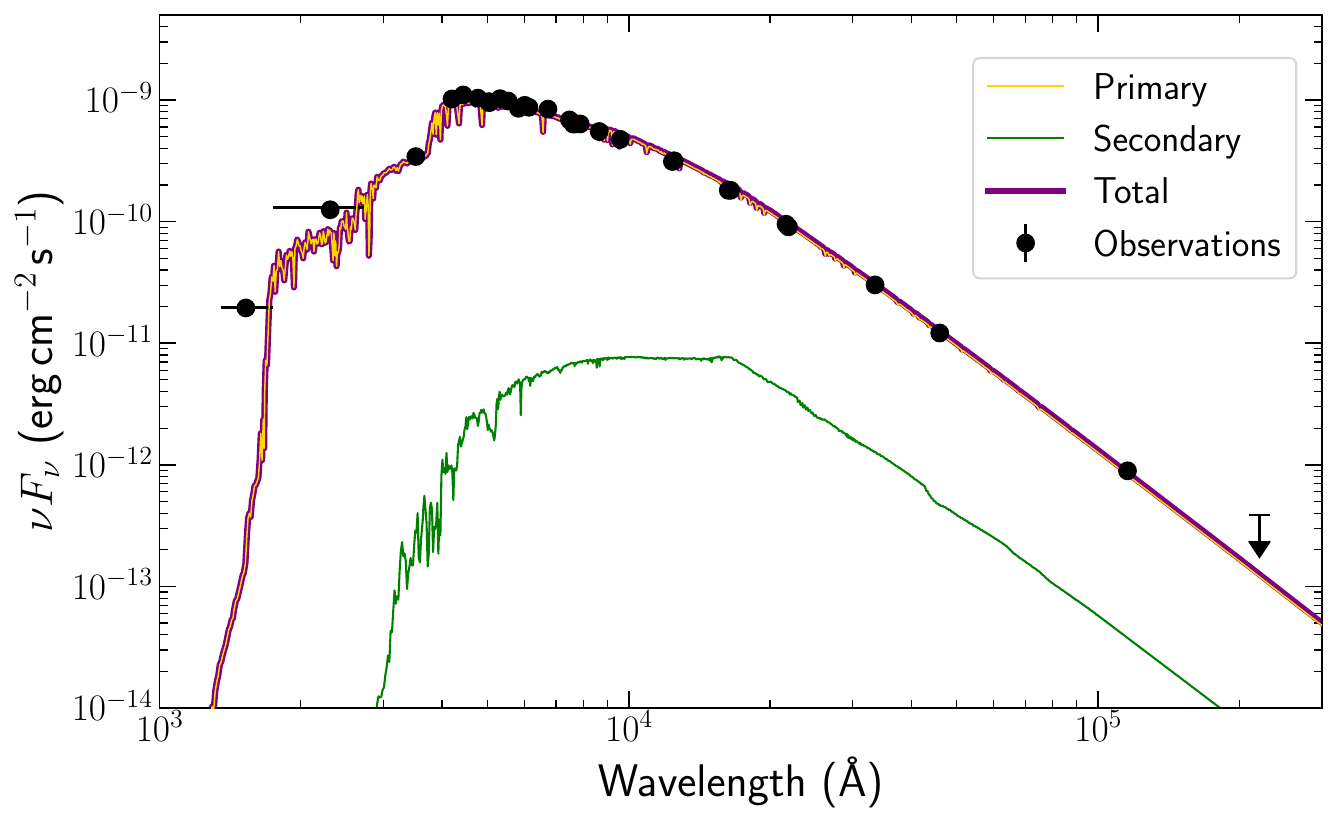}
    \caption{The best-fit SED. The yellow
    and green lines are the best-fit model atmospheres from
    \citet{castelli_kurucz} for the primary and
    secondary, respectively, while the thicker purple line is
    the sum of the two models. At nearly all wavelengths, the
    primary star's emission dominates that of the
    secondary's by at least two orders of magnitude. We have 
    added a horizontal error bar to show the width of the
    GALEX FUV and NUV bandpasses, and an upper limit for the measurement
    in the W4 band. The uncertainties of $\sim$10\% on each flux point
    are too small to be seen at this scale.}
    \label{fig:sed_fit}
\end{figure}

Simultaneously with the light curve fit, we also fit the SED
of the TIC\,435850195 system. The SED fitting technique is 
based upon that presented in \citet{co_cam} and \citet{tic_6332,six_triples}; 
the MCMC implementation itself is derived from that used in 
\citet{jayaraman_5724661}. For a given
trial pair of masses and a system age, we obtain the radius, effective
temperature, and luminosity from the MIST evolutionary tracks. The luminosity 
value is an input into the light curve
calculation, as discussed previously; the radii and effective temperatures
are used to estimate the emitted flux at each wavelength for which we have 
observations, using the \citeauthor{castelli_kurucz} model 
atmospheres.\footnote{We specifically used a model with solar metallicity.}
The best fit SED is shown in Figure \ref{fig:sed_fit}, and the best-fit
stellar parameters are enumerated in Table \ref{tab:fit_system_params}.

\begin{table}
    \centering
    \begin{tabular}{l r }
        \hline
        \hline 
       Parameter  & Value  \\ 
       \hline
        $M_1$* (M$_{\odot}$) & 1.77\,$\pm$0.08 \\
        $M_2$* (M$_{\odot}$) & 0.66\,$\pm$0.03 \\
        $R_1$ (R$_{\odot}$) & 2.16$^{+0.11}_{-0.14}$ \\
        $R_2$ (R$_{\odot}$) & 0.62\,$\pm$\,0.02 \\
        $T_{\rm eff, 1}$ (K) & 7520$^{+430}_{-300}$ \\
        $T_{\rm eff, 2}$ (K) & 4250$^{+150}_{-100}$ \\
        $L_1$ ($L_\odot$) & 13.6$^{+2.42}_{-2.11}$ \\
        $L_2$ ($L_\odot$) & 0.11$^{+0.03}_{-0.02}$ \\
        Age* (Myr) & 970$^{+70}_{-160}$ \\
        $i$* ($^{\circ}$) & 73.3$^{+0.6}_{-0.5}$ \\
        $A_V^\dag$ (mag) & 0.42\,$\pm$\,0.02 \\
        Distance$^\dag$ (pc) & 538$_{-64}^{+55}$ \\
        \hline
    \end{tabular}
    \caption{Best-fit parameters from a joint MCMC fit to the SED and light curve.
    Uncertainties are given as 1-$\sigma$ confidence intervals, calculated from
    the posterior distributions shown in Figure \ref{fig:posteriors}.
    Parameters indicated with an * were directly varied
    as part of the MCMC run. The remaining parameters were
    uniquely determined for a given mass and age by its position on
    the MIST evolutionary tracks \citep{mist_i,mist_ii}. The $\dag$
    indicates that we ran one MCMC with the distance and extinction
    as free parameters to ensure
    that our result was sensible, and then fixed them to the respective
    Gaia-determined values for future runs.}
    \label{tab:fit_system_params}
\end{table}

We find that the best-fit parameters yield an SED that agrees
quite well with observations, but slightly underpredicts the two bluest points.
However, we note that the
GALEX FUV and NUV passbands are rather broad (see, e.g., shaded areas in
Fig. 1 from \citealt{galex_mission_paper}, and the horizontal error
bars for the GALEX points in Fig. \ref{fig:sed_fit}).
Our analysis also shows that the secondary contributes
a very small fraction of the overall system light, and is fainter by
at least two orders of magnitude.

\begin{figure} 
    \centering
    \includegraphics[width=\linewidth]{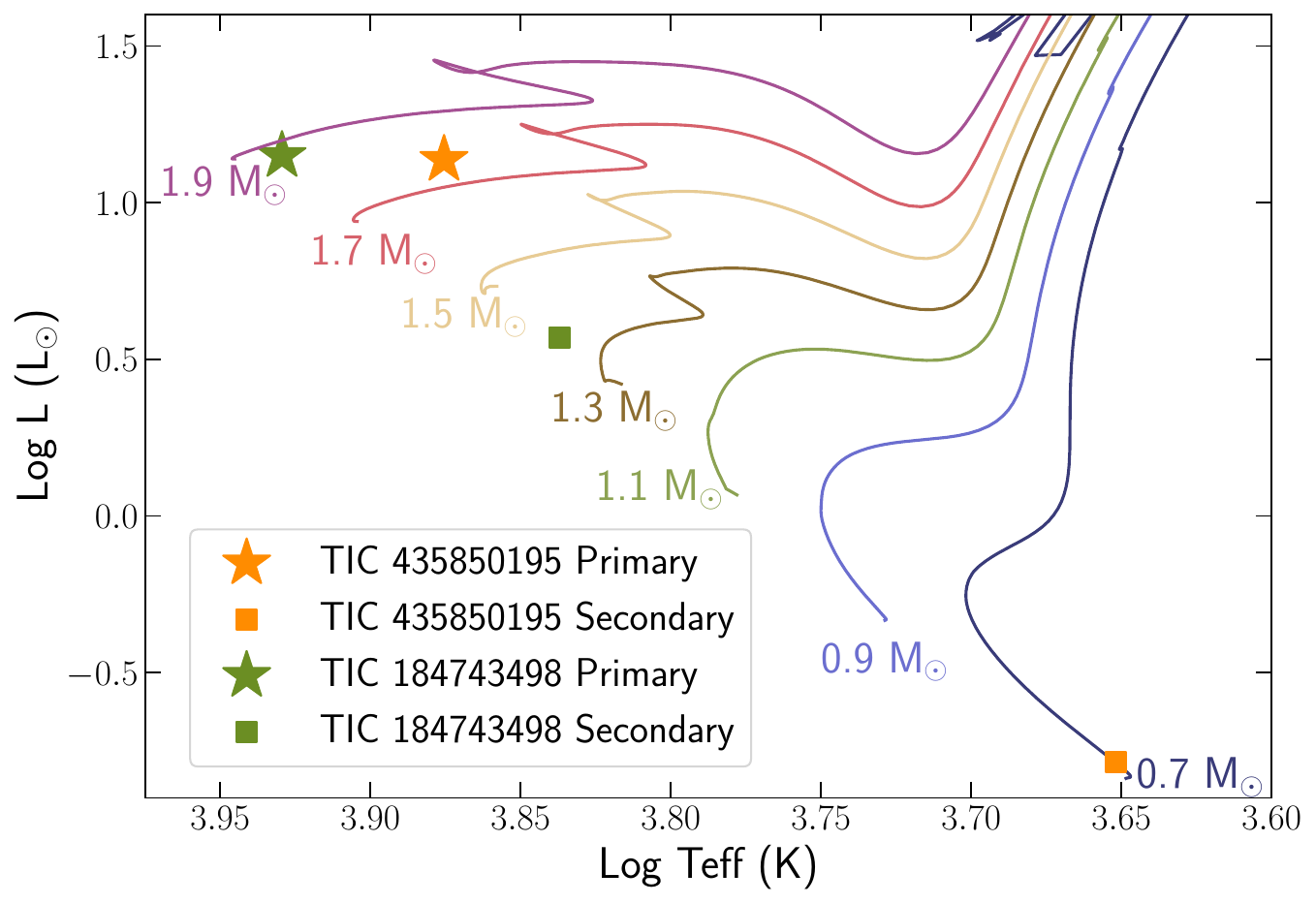}
    \caption{Hertzsprung-Russell diagram showing evolutionary 
    tracks for stars with masses between 0.7 and 1.9 M$_\odot$
    from the MIST tracks \citep{mist_i,mist_ii}. This diagram also shows
    the locations of the primary and secondary components, as
    stars and squares, respectively, for both 
    the TIC\,184743498 (\citetalias{tic_184}) and
    TIC\,435850195 systems.}
    \label{fig:evol_hr_diag}
\end{figure}

\begin{figure*} 
    \centering
    \includegraphics[width=\textwidth]{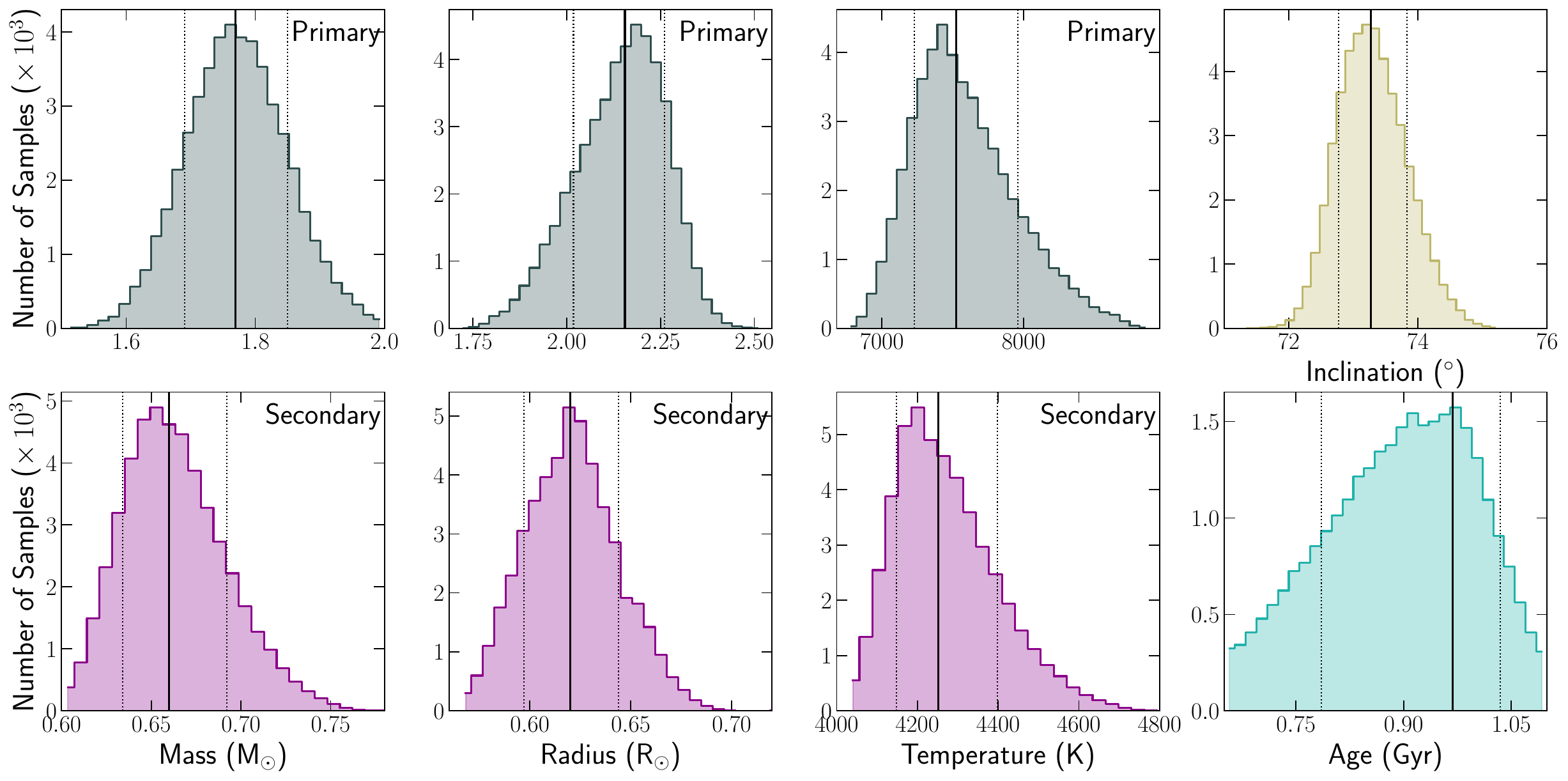}
    \caption{Posterior distributions for stellar parameters
    from the joint MCMC fit to the light curve and the SED. 
    The top panel shows the mass, radius, and temperature
    of the primary (in gray), as well as the inclination 
    (in tan); the bottom panel shows the mass, radius, and
    temperature of the secondary (in magenta), and the
    age of the star (in turquoise). The median value is
    denoted by the solid vertical black lines, while
    1-$\sigma$ confidence intervals are denoted by 
    the dotted black lines. Note that even though we 
    fit for the age indirectly in terms of the ``Equivalent Evolutionary
    Phase'' (EEP), we show the posterior for the age in
    Gyr. Prior to the histogram calculation, 
    the ages were normalized by the density
    of EEP samples in that particular age range.}
    \label{fig:posteriors}
\end{figure*}

\subsection{System Parameter Results}
\label{subsec:sys_params}

The final results of our joint fit to the orbital light curve and the 
composite SED are summarized in Table \ref{tab:fit_system_params}.  
The pulsating primary star has a mass of 1.8\,M$_\odot$ and radius of 2.2\,R$_\odot$, 
and is somewhat evolved off the zero-age main sequence. The secondary is a 
K star of mass 0.66\,M$_\odot$ that contributes under 1\% of the 
system light. The orbital separation is close to 7\,R$_\odot$. Using the 
approximation for the effective Roche lobe (RL) radius $r_L$ from \citet{eggleton_rl}, with a 
mass ratio $q$ of 0.37, we find $r_L$ = 0.59$a$. With $a \simeq 7\,R_\odot$, we 
find that the primary star has $r_L$ = 4.13\,$R_\odot$; thus,
it fills slightly over 50\% of its Roche lobe.  The orbital
inclination angle, a key parameter for this system, is 73.3$^{+0.6}_{-0.5}\,^\circ$.

We note a potentially important discrepancy in what our system parameters 
predict for the $K_1$ velocity of the primary, compared to the measured value
from Gaia, 46.13\,km\,s$^{-1}$. Our best-fit system
parameters indicate a $K_1$ value of $67.41\pm4.23$\,km\,s$^{-1}$, which differs
from the Gaia value by 5.0\,$\sigma$. To decrease this discrepancy to
2-$\sigma$, either the inclination angle must be lowered to be between 
35.6$^\circ$--46.9$^\circ$---which would make the eclipses disappear---or the 
mass of the secondary star must be lowered to
$M_2\lesssim0.5\,$M$_\odot$. Such a star, if near the ZAMS, would
have an insufficient intrinsic luminosity to produce a secondary eclipse
as deep as the one that is observed, which has a depth of approximately
1\%; the eclipses produced for such a low-mass star would have a 
depth of $\sim$0.2\%. We posit that the Gaia uncertainty 
on their measurement of $K_1$ is underestimated by
a factor of 2--3.

Our best fit for the system age suggests that it is approximately 970\,Myr
old. The primary of this system appears to be slightly more evolved
than the one in TIC\,184743498 (the first TAP to be discovered); however,
its mass and other parameters are very similar to the primary in
that system (cf. Table 3 in \citetalias{tic_184}). A
comparison of the evolutionary states of the two systems is shown in Figure 
\ref{fig:evol_hr_diag}, alongside evolutionary tracks for various stellar 
masses, assuming a solar metallicity. Both the TIC 184743498 and TIC 435850195
systems have $R_1/a$ close to 0.3 and, as we will see, have similar 
ranges for their $\delta$\,Scuti pulsations. In TIC 184743498, however, 
the secondary is twice as massive as the one in TIC\,435850195, and there
also likely exists a tertiary component in that system that has 
comparable mass to the secondary.

Posterior distributions for a selection of the parameters 
tracked throughout the MCMC are shown in Figure \ref{fig:posteriors}.

\section{Pulsational Analysis}
\label{sec:puls}

\begin{figure*} 
    \centering
    \includegraphics[width=\textwidth]{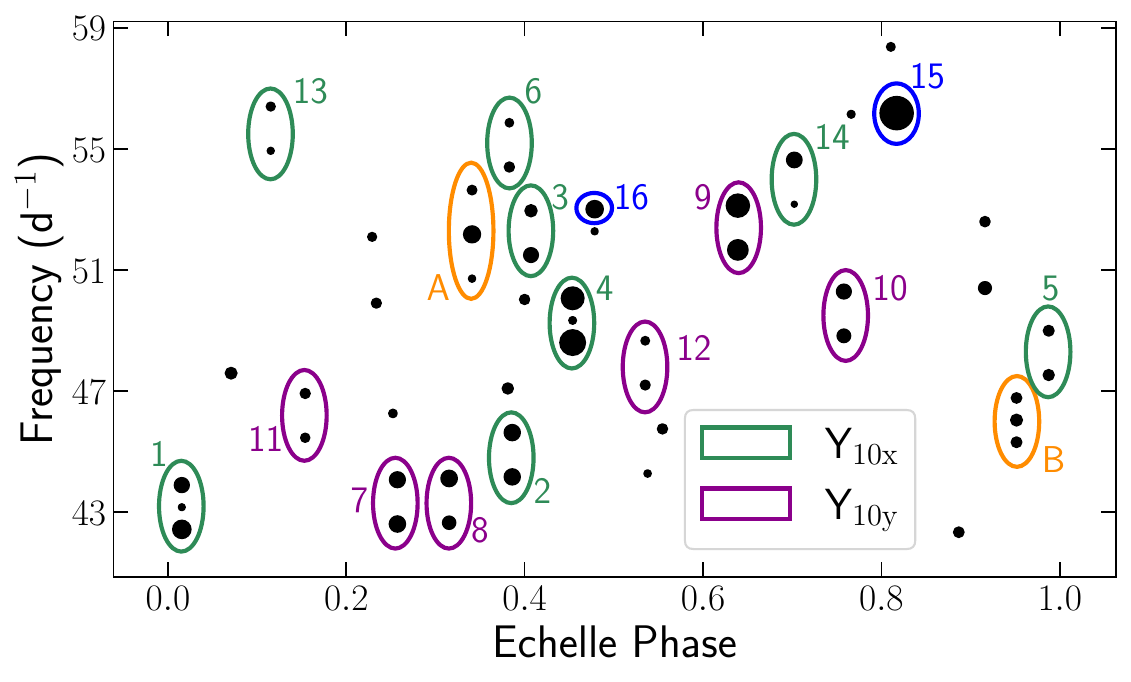}
    \caption{An \'echelle diagram for TIC 435850195, showing the tidally 
    tilted $\delta$\,Scuti pulsations. The abscissa is labeled by the \'echelle
    phase, which is the pulsation frequency $\nu$, modulo the orbital
    period, divided by the orbital period. The sizes of the points are scaled linearly
    according to their amplitudes in the periodogram of the light curve.
    From our analysis, we were able to extract 16 multiplets of tidally 
    tilted pulsations. The doublet dipoles have been circled based on their
    pulsation axes. The numbered dipole doublets are shown in Figures 
    \ref{fig:y10x_ampls} and \ref{fig:y10y_ampls}; the triplets A and B
    are further discussed in Section \ref{subsubsec:triplets}.}
    \label{fig:ech}
\end{figure*}

To analyze the pulsations, we utilized {\tt Period04} 
\citep{period04_lenz_breger} in order to perform a multifrequency
fit for the amplitudes and phases at the time of a reference primary 
eclipse for all the pulsation modes.

We first performed a standard frequency analysis, identifying the most 
significant pulsations. We do this by finding the highest peak in 
the Fourier amplitude spectrum, carrying out a linear least squares 
fit for the amplitude and phase of that frequency, and then 
subtracting it from the light curve. We repeat this process and 
sequentially remove the highest Fourier peaks until the noise floor is reached.
From this initial list of frequencies, we constructed
an \'echelle diagram (Figure \ref{fig:ech}), where the pulsation 
frequency $\nu_{\rm puls}$ is plotted
against the ``\'echelle phase'', which is defined as 
$(\nu_{\rm puls}\,{\rm mod}\,\nu_{\rm orb})/\nu_{\rm orb}$. 

We then proceeded to identify frequencies in this \'echelle diagram that 
could be separated by integer multiples of the orbital frequency. {\tt Period04} 
is able to provide optimal light-curve fits for multiperiodic signals 
including harmonic, combination, and equally spaced frequencies---which is 
essential for the present analysis. We used this option to test whether
or not a given candidate multiplet of frequencies is consistent with 
the assumption of spacings that are related to the orbital frequencies; 
if not, their amplitudes would notably change using the force-fitted 
frequencies. Having identified all possible multiplets, we fixed all 
their frequencies, and calculated the amplitudes and phases for each
pulsation frequency via least-squares fitting of each set of doublet frequencies
to the light curve. All the frequencies in multiplets,
and their amplitudes and phases, are enumerated in Table \ref{tab:freqs}.

If a pair of pulsations share the same
\'echelle phase and are separated in frequency by 2$\nu_{\rm orb}$,
we call that a ``dipole doublet.'' A set of three pulsations
separated by either $\nu_{\rm orb}$ or 2$\nu_{\rm orb}$ (sets
A and B in Figure \ref{fig:ech}) is referred to as either
a dipole or a quadrupole triplet, respectively. 
Dipole doublet modes that have the
same pulsation axis are circled in the same color in Figure \ref{fig:ech}.

\begin{table*}
    \label{tab:freqs}
    \caption{Multifrequency solution for all the observed pulsation multiplets, and
    the two highest-amplitude singlets.
    Multiplets are labeled according to their numbering in the \'echelle
    diagram (Fig. \ref{fig:ech}) and the amplitude-phase reconstructions (Figs. 
    \ref{fig:y10x_ampls} and \ref{fig:y10y_ampls}). We also provide mode 
    identifications for the listed modes; the doublets and singlets are all
    $Y_{\rm 10}$ modes, while the mode identifications for the triplets are
    much more difficult to ascertain. Uncertainties for the frequencies
    range from 0.0002\,d$^{-1}$ for the strongest peaks, to 0.0015\,d$^{-1}$
    for the weakest peaks that are still confidently detected.}
    

    \centering
    \begin{tabular}{c c c c c}
        \hline 
        \hline
        Frequency & Amplitude & Phase & Multiplet & Mode ID \\
        (d$^{-1}$) & (ppt) & (rad) & & \\
        \hline 
        42.4342 & 0.7178\,$\pm$\,0.0433 & 0.595\,$\pm$\,0.060 & 1 & $Y_{\rm 10x}$\\
        43.8971 & 0.4899\,$\pm$\,0.0433 & 0.648\,$\pm$\,0.088 & & \\[.2cm]
        44.1681 & 0.4933\,$\pm$\,0.0436 & 1.008\,$\pm$\,0.088 & 2 & $Y_{\rm 10x}$ \\
        45.6310 & 0.4973\,$\pm$\,0.0436 & 0.933\,$\pm$\,0.088 & & \\[.2cm]

        51.4978 & 0.4858\,$\pm$\,0.0437 & 4.904\,$\pm$\,0.090 & 3 & $Y_{\rm 10x}$\\
        52.9607 & 0.3419\,$\pm$\,0.0437 & 4.383\,$\pm$\,0.128 & & \\[.2cm]

        48.6062 & 1.3657\,$\pm$\,0.0410 & 5.941\,$\pm$\,0.030 & 4 & $Y_{\rm 10x}$\\
        50.0691 & 1.1191\,$\pm$\,0.0410 & 5.800\,$\pm$\,0.037 & & \\[.2cm]
        
        47.5338 & 0.3046\,$\pm$\,0.0439 & 2.494\,$\pm$\,0.144 & 5 & $Y_{\rm 10x}$\\
        48.9967 & 0.2084\,$\pm$\,0.0439 & 2.383\,$\pm$\,0.211 & & \\[.2cm]
        
        54.4057 & 0.2743\,$\pm$\,0.0440 & 0.034\,$\pm$\,0.160 & 6 & $Y_{\rm 10x}$\\
        55.8686 & 0.1446\,$\pm$\,0.0440 & 0.086\,$\pm$\,0.304 & & \\[.2cm]
        
        42.6111 & 0.4958\,$\pm$\,0.0435 & 5.899\,$\pm$\,0.088 & 7 & $Y_{\rm 10y}$ \\
        44.0739 & 0.5621\,$\pm$\,0.0435 & 2.829\,$\pm$\,0.077 & & \\[.2cm]
        
        42.6535 & 0.2938\,$\pm$\,0.0438 & 2.638\,$\pm$\,0.149 & 8 & $Y_{\rm 10y}$ \\
        44.1163 & 0.3933\,$\pm$\,0.0438 & 6.007\,$\pm$\,0.111 & & \\[.2cm]
        
        51.6675 & 0.8787\,$\pm$\,0.0418 & 5.614\,$\pm$\,0.048 & 9 & $Y_{\rm 10y}$ \\
        53.1303 & 1.1997\,$\pm$\,0.0418 & 2.410\,$\pm$\,0.035 & & \\[.2cm]
        
        48.8287 & 0.3127\,$\pm$\,0.0437 & 4.141\,$\pm$\,0.140 & 10 & $Y_{\rm 10y}$ \\
        50.2915 & 0.4818\,$\pm$\,0.0437 & 0.840\,$\pm$\,0.091 & & \\[.2cm]
        
        45.4612 & 0.1319\,$\pm$\,0.0440 & 4.629\,$\pm$\,0.333 & 11 & $Y_{\rm 10y}$ \\
        46.9240 & 0.2208\,$\pm$\,0.0440 & 1.630\,$\pm$\,0.199 & & \\[.2cm]
        
        47.2029 & 0.1807\,$\pm$\,0.0439 & 2.533\,$\pm$\,0.243 & 12 &  $Y_{\rm 10y}$\\
        48.6658 & 0.3510\,$\pm$\,0.0439 & 4.766\,$\pm$\,0.125 & & \\[.2cm]

        54.9415 & 0.0786\,$\pm$\,0.0440 & 0.541\,$\pm$\,0.561 & 13 & Y$_{\rm 10x}$ \\
        56.4043 & 0.1145\,$\pm$\,0.0440 & 0.840\,$\pm$\,0.385 & & \\[.2cm]

        53.1766 & 0.2935\,$\pm$\,0.0437 & 1.661\,$\pm$\,0.149 & 14 &  Y$_{\rm 10x}$\\
        54.6395 & 0.5216\,$\pm$\,0.0437 & 0.103\,$\pm$\,0.084 & & \\[.5cm]
        
        56.1862 &2.405\,$\pm$\,0.038 & 4.186\,$\pm$\,0.016 & 15 & likely $Y_{\rm 10z}$ \\[.2cm]
        
        53.0128 & 0.7243\,$\pm$\,0.0435 & 2.042\,$\pm$\,0.060 & 16 & likely $Y_{\rm 10z}$ \\[.5cm]

        50.7180 & 0.1061\,$\pm$\,0.0436 & 5.180\,$\pm$\,0.411 & A & Unclear \\
        52.1809 & 0.6511\,$\pm$\,0.0436 & 2.161\,$\pm$\,0.067 & & \\
        53.6437 & 0.1766\,$\pm$\,0.0436 & 5.450\,$\pm$\,0.247\\[.2cm]

        45.3131 & 0.2305\,$\pm$\,0.0439 & 0.475\,$\pm$\,0.190 & B & $Y_{\rm 10}$? \\
        46.0446 & 0.2595\,$\pm$\,0.0439 & 2.875\,$\pm$\,0.169 & & ($I_p=30^\circ$) \\
        46.7760 & 0.2453\,$\pm$\,0.0439 & 0.236\,$\pm$\,0.179 \\
        \hline
    \end{tabular}
\end{table*}

\subsection{Dipoles and Tri-Axial Pulsations}
\label{subsec:doublets}
\begin{figure*} 
    \centering
    \includegraphics[width=\textwidth]{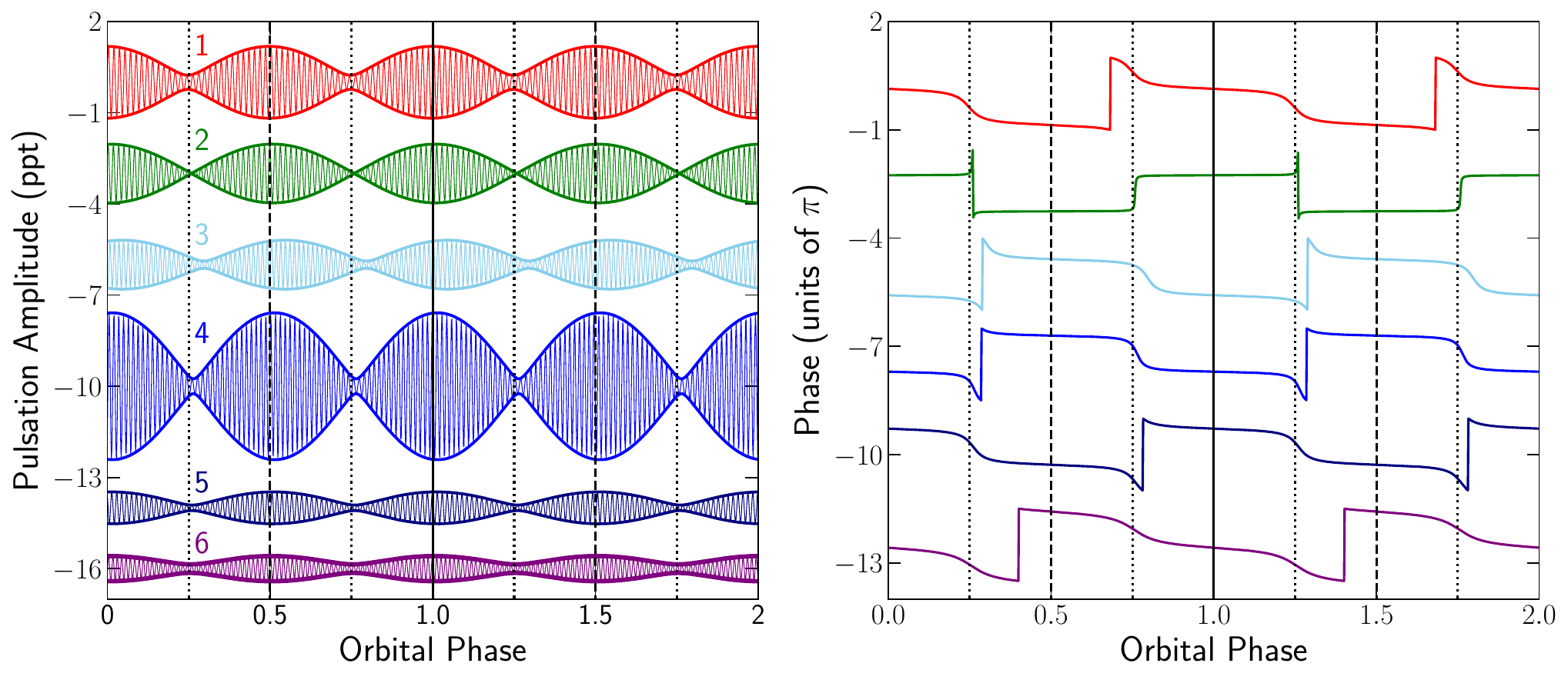}
    \caption{The amplitude (left) and phase (right) variations of the six
    strongest $Y_{\rm 10x}$ modes in TIC 435850195, as a function of
    orbital phase. These modes exhibit
    amplitude maxima at primary eclipse (phase 0, 1, and 2), and
    amplitude minima at the maxima of the ELVs
    (phases 0.25, 0.75, 1.25, and 1.75). These modes also undergo a $\pi$ phase shift at 
    minimum amplitude. The abrupt 2$\pi$ jumps in phase are 
    an artifact of the cyclic nature of the phases and
    are not physically meaningful.}
    \label{fig:y10x_ampls}
\end{figure*}

\begin{figure*} 
    \centering
    \includegraphics[width=\textwidth]{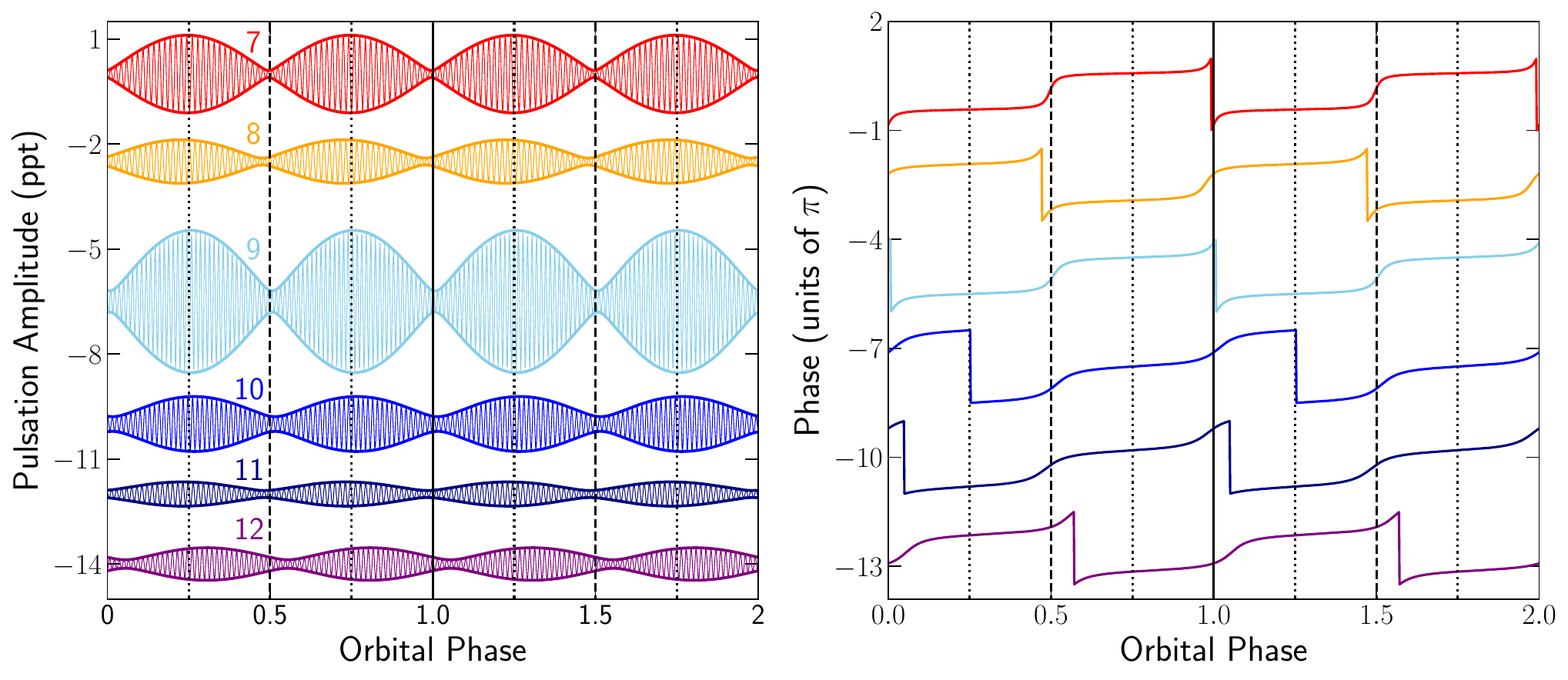}
    \caption{As Figure \ref{fig:y10x_ampls}, but for the six
    highest-amplitude $Y_{\rm 10y}$ modes in TIC 435850195. These modes have
    amplitude minima at primary eclipse (phase 0, 1, and 2), and
    amplitude maxima at the ELV maxima (phases 0.25, 0.75, 1.25, and 1.75). 
    These modes also undergo a $\pi$ phase shift at minimum amplitude.
    In every way, they are identical to the $Y_{\rm 10x}$ modes, 
    except for being shifted by 90$^\circ$ in orbital phase.
    The phases for mode 12 have uncertainties of 0.243 and 0.125,
    respectively, which may partially explain
    why the calculated pulsation amplitude minima and maxima are slightly
    offset from the times of eclipse and ELV maxima.}
    \label{fig:y10y_ampls}
\end{figure*}

To identify the $\ell$ and $m$ values for a given pulsation, we 
calculated its amplitude and phase as a function of the orbital phase. This was done
following the formalism from \citet{hd_265435} for the
special case of a doublet. The amplitude $C$ and phase 
$\phi_{\rm mult}$ of a doublet, as a function of
orbital phase $\Phi$, can be written as
\begin{eqnarray}
    C^2(\Phi) &= A^2 + B^2 + 2AB\cos(\phi_A - \phi_B - 2\Phi)\\
    \phi_{\rm mult} &= {\rm ATan2}\left[\frac{A\sin(\phi_A - \Phi) + B\cos(\phi_B + \Phi)}{A\cos(\phi_A - \Phi) + B\cos(\phi_B + \Phi)}\right]
\end{eqnarray}
Here, $A$ and $B$ are the amplitudes of the pulsations that
comprise the multiplet, and $\phi_A$ and $\phi_B$ are their
phases at a particular epoch time $t_0$,
which we have defined as the time of the first primary eclipse 
in the data (see Section \ref{sec:obs}). 
These amplitude and phase variations are shown as a 
function of orbital phase in Figures
\ref{fig:y10x_ampls} and \ref{fig:y10y_ampls}. 

The doublets with an amplitude maximum at the time of primary 
eclipse exhibit $\pi$ phase shifts at orbital phases 
0.25 and 0.75 (i.e., times of ELV maxima). These correspond to modes 1--6, 
as well as modes 13 and 14, in Figure \ref{fig:ech}. 
We identify these as $Y_{10x}$, modes following the notation in 
\citetalias{tic_184}, {where $\ell=1$, $m=0$, and the $x$-axis 
corresponds to the tidal axis.} In this case, we are looking toward the 
pulsation poles of these modes during eclipse, and the phase 
is expected to flip by $\pi$ radians when we, the observers, 
switch to viewing the other pulsational hemisphere at orbital quadrature.   
These modes are similar to the ones presented in Figure 2 of \citealt{reed_tipped_2005}
(henceforth shortened to \citetalias{reed_tipped_2005}).\footnote{Their
notation $I_r$ corresponds to the inclination angle, which we 
refer to using $i$. We will use $i$ throughout to avoid confusion.}
Their predictions for $i = 75^\circ$, and $I_p=90^\circ$---when 
the pulsation axis is pulled fully into alignment with the line of 
apsides with the binary---agree with our observations for these modes.

In contrast, the other dominant set of modes from the \'echelle,
numbered 7--12, exhibit amplitude minima at the times of 
primary and secondary eclipse, and $\pi$ phase shifts at these
times. Thus, these modes exhibit exactly the same behavior 
as the $Y_{\rm 10x}$ modes, except they are shifted by 
90$^\circ$ in orbital phase. This similarity with the 
$Y_{10x}$ modes extends to the fact that they are also doublets 
that show no significant central peak. As was done in \citetalias{tic_184},
we tentatively identify these pulsations with $Y_{10y}$ modes.  
The description of what the observer sees as a function of orbital 
phase is exactly the same as for the $Y_{10x}$ modes, except
shifted by 90$^\circ$ in orbital phase. In other words, 
this is a $Y_{10}$ mode with a pulsation axis lying along 
the system's $y$ axis, perpendicular to the axes of both the
tidal bulge and the orbital momentum.

If we were to adopt the formalism of 
\citetalias{reed_tipped_2005}, the amplitude variations of these
modes, at first glance,
appear most similar to $Y_{\rm 11x}$ modes (cf. their
Figure 2.2, with $i=75^\circ$, and $I_p=90^\circ$). However,
we highlight that the corresponding \citetalias{reed_tipped_2005} periodogram shows a 
detectable central peak that we do not observe in these modes of TIC 435850195. 
We suggest that the explanation given in Section 6 of
\citetalias{tic_184} reveals the true provenance of these 
modes, and identify these as $Y_{\rm 10y}$ modes.

\begin{figure*} 
    \centering
    \includegraphics[width=\textwidth]{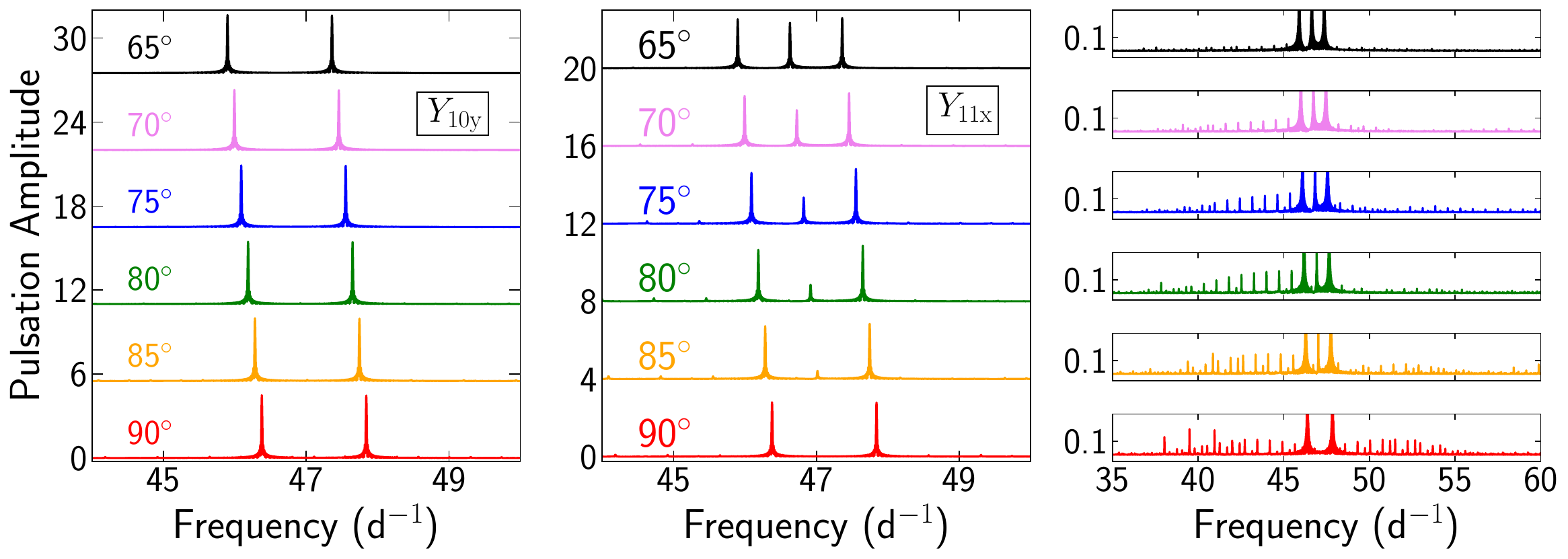}
    \caption{Simulated {$Y_{\rm 10y}$ (left panel)} and 
    $Y_{\rm 11x}$ modes (middle panel) for a star 
    similar to TIC\,435850195. These show that at the
    inclination angle of the system, roughly
    75$^\circ$ (indicated in blue), we expect to see a central
    peak that is approximately half the amplitude of the
    sidelobes for any $Y_{\rm 11x}$ modes that may be present. 
    {In contrast, we observe only a frequency doublet of
    pulsations, regardless of inclination angle, for a $Y_{\rm 10y}$ 
    mode.}
    The right panel shows a vertical zoom in on the periodogram {
    for the $Y_{\rm 11x}$ modes}
    to highlight the Fourier peaks arising from
    eclipse mapping (also called spatial filtration),
    further discussed in Section \ref{subsec:astero_constraints}.
    Note the differing x-axis scales between the middle and right panels.}
    \label{fig:y11x_sim}
\end{figure*}

To further disprove the hypothesis that the observed 
doublets 7-12 arise from $Y_{\rm 11x}$
modes, we simulated the frequency spectrum expected from these modes
as a function of orbital inclination angle $i$ (shown in Figure 
\ref{fig:y11x_sim}). This figure, similar to Figure 3 from 
\citetalias{reed_tipped_2005}, shows that $Y_{\rm 11}$ modes must
show a significant central peak for a binary with $i\lesssim 75^\circ$.
 However, there exists no evidence
for such a central peak in any of the doublets whose amplitude
and phase variations are shown in Figure
\ref{fig:y10y_ampls}. 

As part of our verification procedure to confirm that the
doublets do not arise from $Y_{\rm 11x}$ modes, we also 
calculated the ratios of the central peak to the sidelobes
that would be expected for these modes, as a function of 
inclination angle $i$.
This calculation was based on the peaks 
shown in Fig.~\ref{fig:y11x_sim}. In order to estimate the noise level,
we fit a Rayleigh distribution
to the white noise in the periodogram between 60 and 80\,d$^{-1}$,
which is a region without any significantly-detected frequencies. The 97\% 
upper limit for this distribution, which has a mode of 18.74 ppm, was 50 ppm;
below this value, we are unable to claim a detection of any
periodicity in the light curve, even at marginal significance.

\begin{figure}
    \centering
    \includegraphics[width=\linewidth]{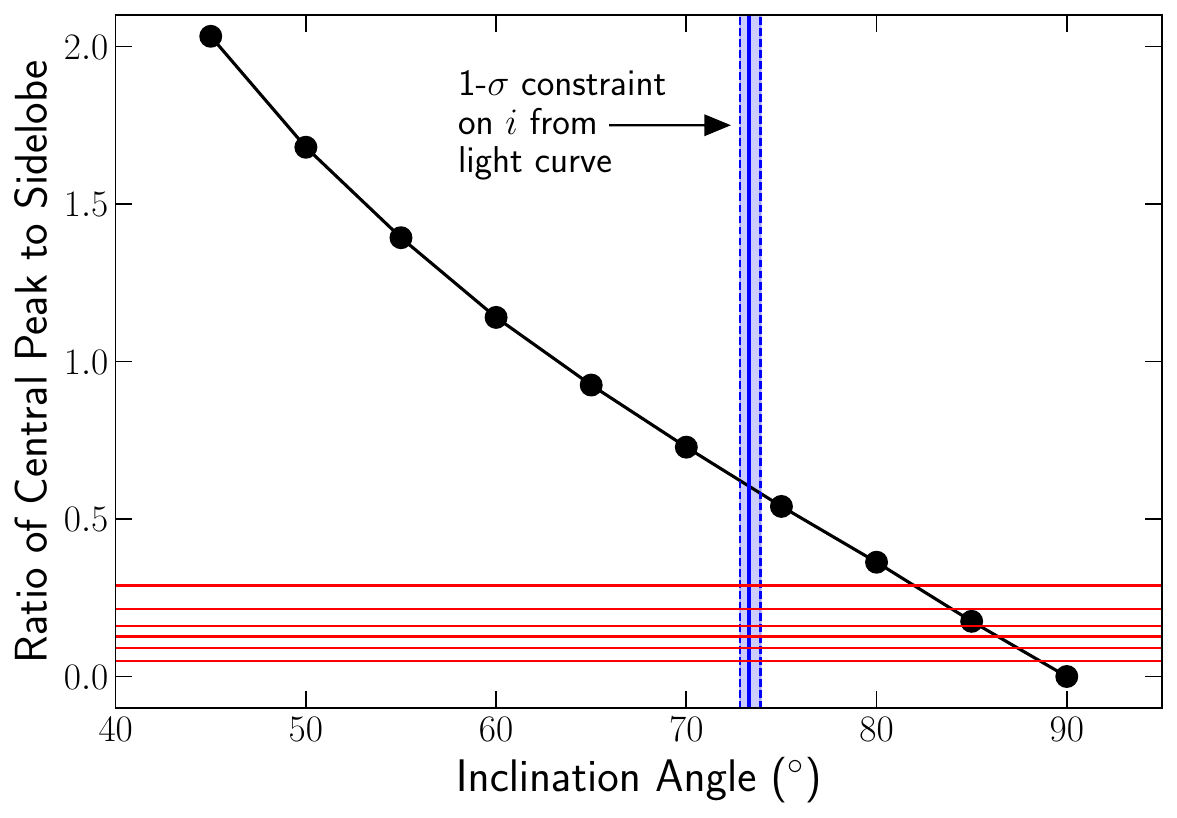}
    \caption{A comparison of the ratios of the central peak to
    the sidelobe amplitudes for $Y_{\rm 11x}$ modes to the limits
    obtained from the $Y_{\rm 10y}$ modes in TIC 435850195. The
    red lines correspond to the ratio of the 2-$\sigma$ detection
    threshold to the mean amplitude of each $Y_{\rm 10y}$ doublet.
    The blue shaded area corresponds to the 1-$\sigma$ uncertainty
    for the inclination angle of the system, from our light curve fit.}
    \label{fig:y11x_y10y_limits}
\end{figure}

Using this value,
we set empirical upper limits on the presence of
any central peak for all of modes 7--12, and 
thereby establish secure upper limits on the ratio of
any putative central peak to its sidelobes (shown in 
Figure \ref{fig:y11x_y10y_limits}). For an inclination angle of
$\sim$\,75$^\circ$, we find that the ratio of the central 
peak to the sidelobe should be slightly over 0.5; the limit 
for the observed modes, at the observed inclination angle, 
is under 0.4 for two of the modes, and 
under 0.1 for three of the modes. This further reinforces the 
fact that the modes we are seeing are, in fact, not $Y_{\rm 11x}$ modes,
and instead are the novel $Y_{\rm 10y}$ modes discovered and 
reported by \citetalias{tic_184}.

\subsection{Other Observed Pulsations}
Figure \ref{fig:ech} shows a number of other pulsation modes in addition to the
dipole doublets discussed previously; these include singlets,
as well as two potential triplets---one a dipole (B), and the other a quadrupole 
mode (A). Most of the pulsations that we see in TIC 435850195 can be
explained within the tidal tilting framework, including
some that may only be partially tilted (``tidally perturbed''). For the
purposes of our analysis, we focus on the highest-amplitude singlets,
and those modes that are in clear multiplets separated by
multiples of $\nu_{\rm orb}$.

\subsubsection{On the Existence of Radial Modes}

The \'echelle diagram from Figure \ref{fig:ech} shows mostly 
dipoles and one quadrupole mode, but there are two prominent pulsation
singlets: one\footnote{This mode
has a low-significance sidelobe separated from it by $\nu_{\rm orb}$, but
it does not have a significant effect on the overall amplitude and phase
variability of this pulsation mode, so we choose to interpret it as a 
singlet rather than as a flavor of $Y_{\rm 10}$ mode.}
at 53.01285 d$^{-1}$ (\#16), and another one at 
56.18624\,d$^{-1}$ (\#15). At first glance, these may appear to be 
radial modes. However, this interpretation is most likely incorrect.

We first calculate the period ratio of these two oscillations,
along with the large frequency spacing, to evaluate this
hypothesis. Using the models of \citet{stellingwerf_ratios},
we find that the ratios of the periods of the first, second, 
and third overtones to the fundamental period are 0.772, 0.629, 
and 0.527, respectively. The period ratio of these two modes is 
0.944, so it is unlikely to be a low-order radial mode. For
higher-order radial modes, the asymptotic (large) frequency 
spacing is given by
\begin{equation}
    \label{eq:deltanu}
    \Delta\nu = \Delta\nu_\odot \sqrt{\frac{\bar{\rho}}{\bar{\rho}_\odot}}.
\end{equation}
Here, $\Delta\nu_\odot$ is a constant, 135\,$\mu$Hz, and $\bar{\rho}$ is the
mean stellar density. Utilizing our mass and radius values from Table
\ref{tab:fit_system_params} yields 
a stellar density of 0.249\,g\,cm$^{-3}$. Substituting this 
into Equation \ref{eq:deltanu} yields 
$\Delta\nu \approx 57$\,$\mu$Hz, or 4.9\,d$^{-1}$, which is
larger than the separation between these two singlet frequencies.
Consequently, radial modes probably cannot explain these two observed singlets. 

These pulsations, being singlet modes, do not 
exhibit the amplitude and phase variability that we observe
in the $Y_{\rm 10x}$ and $Y_{\rm 10y}$ modes. We thus interpret
at least one of these modes, or possibly both, as non-radial 
$Y_{\rm 10z}$ modes, similar to the 
framework presented in \citetalias{tic_184}, 
making TIC\,435850195 a novel tri-axial pulsator. We also suggest
the possibility that one of these
modes (but not both) could be a radial ($Y_0$) mode instead of
 $Y_{\rm 10z}$.

\subsubsection{Triplets}
\label{subsubsec:triplets}
Triplet A (as numbered in Figure \ref{fig:ech}) has each of its components
separated by $2\nu_{\rm orb}$ (making it a quadrupole mode), while triplet B 
has a separation of just $\nu_{\rm orb}$, making it a dipole mode. 
Using the formalism from \citet{hd_265435}
to reconstruct the amplitude and phases of these triplets, 
as we did for the other modes in Subsection \ref{subsec:doublets}, we
find that triplet A has two amplitude maxima per orbit; these
exactly correspond to the ELV maxima. However, there are no $\pi$ phase
jumps, as would be expected for an $m=0$ mode {whose pulsation
axis is the $z$-axis}; moreover, given that
the components are separated by 2$\nu_{\rm orb}$, this is
likely to be a $\ell = 2$ mode. While there is not an exact
match from \citetalias{reed_tipped_2005} for a quadrupole mode with
a strong central peak and sidelobes separated by 2$\nu_{\rm orb}$, 
we note that there could be additional multiplet components that 
are obscured by the noise, making a conclusive mode identification 
difficult with the existing data set.

There are two other possibilities that we could explore for these modes:
Either it is an $\ell=2$ mode, with $m=\pm2$, or it is
rotationally split; this latter hypothesis will be discussed
further in Section \ref{subsubsec:no_rot_split}. 
We note that \citetalias{reed_tipped_2005} indicate
that for $(\ell,m) = (2, \pm2)$, there should be no $\pi$ phase jumps
under any condition. However, 
our inclination angle of $i\sim73.5^\circ$ should yield a quintuplet
spaced by $\nu_{\rm orb}$ for a fully tidally tilted mode ($I_p=90^\circ$).
There are no clear matches from \citetalias{reed_tipped_2005} for
this scenario, even for tidally perturbed modes ($I_p < 90^\circ$).
We also note that it is possible for $\ell=2$ modes with different
$m = -2, 0, 2$ to couple and produce what we refer to as 
$Y_{2+}$ and $Y_{2-}$ modes; however, these modes are predicted
to have two $\pi$ phase jumps for each orbital cycle, which we 
do not observe here. Excluding these possibilities suggests that this
triplet is difficult to interpret with our current understanding
of tidally tilted pulsations.

Triplet B has its components separated by $\nu_{\rm orb}$; each
component frequency has a comparable amplitude. 
This case could correspond to the $I_p=30^\circ$
case for a $Y_{\rm 10}$ mode, making this a tidally perturbed mode
rather than a fully tilted mode. Given that the separation between the
triplet components is exactly $\nu_{\rm orb}$, we
rule out the rotational splitting explanation (see Section \ref{subsubsec:no_rot_split}).
Another possibility is that this pulsation is tidally
perturbed as part of a higher-order mode (e.g., with $\ell=2$), 
with the other frequencies being indistinguishable from the 
noise (cf. the $I_p=30^\circ$ case, for 
$i=75^\circ$ in Fig. 3.4 in \citetalias{reed_tipped_2005}).
This triplet could also arise from the 
coupling of modes with different $\ell$ values; however,
that calculation is beyond the scope of this work, and will be
further explored in Fuller et al. (in prep).

We note that neither triplet is likely to be a $Y_{\rm 11x}$ mode,
based upon the ratio of the amplitudes of the peaks shown in the 
simulated periodograms in Figure \ref{fig:y11x_sim}. Specifically,
the central peak is the strongest component of the multiplet, when
for $Y_{\rm 11x}$ modes, it is usually the weakest, when observed
at $i\sim75^\circ$.

\subsubsection{Ruling Out Rotational Splitting}
\label{subsubsec:no_rot_split}
Rapidly-rotating stars can have their pulsations split by the 
rotational frequency, yielding multiplets whose
components are given by the following expression:
\begin{equation}
    \nu_{n,\ell,m} = \nu_{n,\ell} + m(1-C_{n,\ell})\nu_{\rm orb},
\end{equation}
where $C_{n,\ell}$ is the Ledoux constant. For the observed 14 doublets 
that we identified as $Y_{\rm 10x}$ and $Y_{\rm 10y}$ modes, 
there are two compelling arguments against these arising from 
rotational splitting. First, the splitting of the doublets is equal 
to 2$\nu_{\rm orb}$ to within an rms fractional uncertainty 
of $\lesssim 0.001$.  Therefore, the Ledoux coefficient would have to 
be $\lesssim 0.001$, which is implausible---for $\delta$\,Scuti stars,
this value is between 0 and 0.2 \citep{goupil_dsct}, with a specific
value of 0.08 for the $\delta$\,Scuti star KOI-976 \citep{ahlers_ledoux_value}.
Second, the phase shifts between the doublet elements in the 
rotational-splitting scenario can, in principle, take on any value, 
including $\pi$.  However, the likelihood that all of the doublets 
would have $\pi$ phase shifts between their elements at exactly
the same orbital phase is very low, if not zero.  Thus, rotational 
splitting is categorically not the origin of the observed doublets.

For triplet A, despite the difficulty in conclusively assigning a 
mode to it, we suggested that it may be an 
$|m| = 2$ mode. As a result, $C_{n,\ell}$ must again approach 0 to within 
one part per thousand in order to explain this mode as
rotational splitting. A similar argument for $m=1$ holds
for triplet B. As a result, we can eliminate the
rotational splitting hypothesis as a viable explanation for
what we observe in this star's pulsation spectrum.

\section{Discussion}
\label{sec:disc}

We have conclusively shown that TIC 435850195 pulsates along three
different orthogonal axes. The system exhibits almost exclusively
dipole modes, with one likely quadrupole 
mode. In this section, we introduce a toy model for mode coupling
in the TIC 435850195 system, and further discuss the multiplets and 
the effects of eclipse mapping on the
observed periodogram. 

\subsection{Toy Model for Tidal Tilting}

As discussed in \citetalias{tic_184} and in greater detail by Fuller
et al. (in prep), the perturbation $T$ to the potential from the 
tidal bulge in a close binary can be represented by
\begin{equation}
T \propto \frac{x^2}{r^2},
\end{equation}
where $\hat{x}$ is the direction along the tidal bulge, $\hat{z}$ 
is the orbital angular momentum axis, and $\hat{y}$ is a direction 
perpendicular to both.  In turn, this can be written in terms of 
spherical harmonics as
\begin{equation}
T \propto Y_{22,z} +\sqrt{\frac{2}{3}}\, Y_{20,z}+Y_{2-2,z}.
\label{eqn:tidal_perturbation}
\end{equation}
Here, the $z$ subscript indicates that angles in these spherical
harmonics are measured with respect to the $z$ axis. As discussed 
briefly in \citetalias{tic_184} and originally in \citet{ttp_modeling},
a perturbation analysis suggests possible  mode couplings that
the tidal bulge can induce.  Equation (\ref{eqn:tidal_perturbation})
indicates that the $Y_{22,z}$ term in the tidal perturbation
can couple the mode $Y_{11,z}$ with $Y_{1-1,z}$ 
(with $\Delta m = +2$), and vice versa, due to the 
$Y_{2-2,z}$ part of the tidal perturbation. The $Y_{20,z}$ 
term in the tidal perturbation couples to none of the 
$\ell = 1$ modes.  Since the $\ell = 1$ modes represent the vast majority of
what we detect in TIC 435850195, we do not consider the
coupling of $\ell=2$ modes---which will be discussed in 
Fuller et al. (in prep).

The  tidal perturbation thus gives new eigenmodes which are the 
sum and difference of the normal $\ell = 1$ modes.  Here, we 
compare the unperturbed dipole modes ($D_{1\pm1}$) to 
the tidally perturbed ones ($D_{\pm}$):
\begin{eqnarray*}
D_{1\pm1} \propto {\rm Re}\left\{Y_{1\pm 1,z} \,e^{i \omega_1 t} \right\} & \propto ~\sin \theta \cos(\omega_1t \pm \phi) \\
D_{+} \propto {\rm Re} \left\{ \left(Y_{11,z} + Y_{1-1,z} \right) e^{i \omega_+ t} \right\} & \propto ~\sin \theta \sin \phi \sin \omega_+t \nonumber \\
D_{-} \propto {\rm Re} \left\{ \left(Y_{11,z} - Y_{1-1,z} \right) e^{i \omega_- t} \right\} & \propto ~\sin \theta \cos \phi \cos \omega_-t \nonumber
\end{eqnarray*}
Here, $\omega_+$ and $\omega_-$ are the two perturbed eigenfrequencies, and Re\{\} 
represents the real part of the complex exponentials.  The fundamental 
difference between the unperturbed $\ell=1$ modes and the new perturbed 
modes is that the unperturbed modes are waves that travel around the 
equator of the star, while the perturbed modes are standing waves, 
much like a $Y_{10}$ mode. In fact, the perturbed modes can be 
written explicitly as ``tidally tilted'' modes with axes 
along the $x$ and $y$ axes:
\begin{eqnarray}
D_{+} & \propto ~y \sin \omega_+t ~\propto ~ Y_{10y} \sin \omega_+t \\
D_{-}  & \propto ~x \cos \omega_-t ~\propto ~Y_{10x} \cos \omega_-t 
\end{eqnarray}
These are exactly $Y_{10}$ modes with pulsation axes along 
the $y$ and $x$ directions, respectively. They describe the 14 
tidally tilted dipole doublet modes we have detected in TIC 435850195, shown
in Figures \ref{fig:ech}, \ref{fig:y10x_ampls}, and \ref{fig:y10y_ampls}.

\subsection{Other Possible Multiplet Groupings}

We note that the interpretation of certain groupings of peaks in 
the \'echelle (Fig. \ref{fig:ech}) can be subjective. Specifically,
we discuss mode 16, which we choose to interpret as a singlet,
and modes 1 and 4, which we choose to interpret as a doublet $Y_{\rm 10x}$
mode, even though there is a very weak central component present.

First, we discuss the alignment of the frequency doublet (50.407261, 52.60155). This
is split by $3\nu_{\rm orb}$, which does not correspond to any clear
mode identification in the tidal tilting framework. Comparison with
the modes from \citetalias{reed_tipped_2005} also does not yield a 
clear match for two significant peaks separated by $3\nu_{\rm orb}$.
This could either be an octupole mode, with the third component
(also spaced by $3\nu_{\rm orb}$) hidden below the noise. An
alternative (and perhaps more likely) explanation is that this 
is an $\ell=3$ mode, which
might be able to produce doublets that are spaced by $3\nu_{\rm orb}$
in a triaxial tidally tilted
pulsator (Fuller et al. in prep).
We also note that high-$\ell$ modes are often strongly geometrically canceled
\citep{dziembowski_geom_cancel}, which could make this a chance alignment.

What we have tentatively called a singlet, Mode 16, also has another companion
peak with a much smaller amplitude separated by $-\nu_{\rm orb}$.
However, we note that such a small peak does not induce any significant
amplitude and phase variability in this mode; rather, this could be part of a
triplet, with the peak at $+\nu_{\rm orb}$ buried in the noise. In that 
case, this could plausibly be interpreted as a dipole triplet. However,
we would need data with a lower noise floor in order to extract this frequency
and conclusively assign a mode identification to it, using the amplitude
and phase variations throughout the orbit. 

We opted to interpret modes 1 and 4 as doublet $Y_{\rm 10x}$ modes, despite
the presence of a weak ($\lesssim$\,10--20\%) central peak in both of them. Comparing these to
Figure 3 of \citetalias{reed_tipped_2005} suggests that these, instead of
being fully tidally tilted pulsations ($I_p = 90^\circ$), may actually 
be tidally perturbed, or only partially tilted $Y_{\rm 10x}$ modes. 
The presence of the weak central peak is a close match for the 
$I_p = 60^\circ$ and $I_p = 75^\circ$ cases, suggesting that these
modes may not have their axis fully tilted into the orbital plane. 
If this were true, this star would have a unique combination of tidally perturbed
modes, tidally tilted modes, and non-tidally tilted (singlet) pulsation modes,
making the pulsation behavior far richer than initially thought.

\subsection{Further Asteroseismic Constraints}
\label{subsec:astero_constraints}

This star exhibits 14 dipole doublet pulsations, 2 singlet pulsations, and
2 triplet pulsations---one dipole, and one quadrupole. We suggest that the
triplet dipole mode may not be fully tidally tilted, and that the quadrupole
mode is difficult to interpret as part of the tidal tilting framework. 
Some of these modes may be related to the coupling of $\ell=2$ modes,
or coupling between modes that have different $\ell$ values---a derivation
of which is beyond the scope of this work, and will be addressed by 
Fuller et al. (in prep). These modes may also be strongly affected
by the Coriolis force in the star, which has been neglected in our
modeling of previously-discovered TTPs.

An explanation that has sometimes been posited for the 
multiplets that are observed in \'echelle diagrams is 
``eclipse mapping,'' or ``spatial filtration''
(see, e.g., \citealt{u_gru_perturbed}, and references therein,
including \citealt{spatial_filt}). This phenomenon occurs when 
the occultation of the pulsating star obscures different parts
of the surface of a star that is undergoing non-radial pulsations.
This leads to considerably complicated variations in the amplitudes 
of pulsation multiplets (see, e.g., Fig. 10 of
\citealt{u_gru_perturbed}). However, we note that the periodogram
of pulsations that have been spatially filtered will exhibit
a long series of peaks split by $\nu_{\rm orb}$ that grow more
pronounced as the inclination angle increases, as seen in the
right panel of Figure \ref{fig:y11x_sim}.
This effect is most pronounced for binary systems with $i\sim90^\circ$,
and becomes nearly imperceptible (given the noise
properties of the data) for $i\sim75^\circ$. Even the strongest
peaks arising from eclipse mapping are only $\lesssim 5$\% of
the $Y_{\rm 11x}$ triplet.
Thus, we can be confident that the observed multiplets are
indeed caused by tidal phenomena, and are not simply a function of 
our observational perspective on the star.

\section{Conclusions}

In this work, we report the discovery of the second ever
tri-axial pulsator, exhibiting $Y_{\rm 10x}$, $Y_{\rm 10y}$, and
$Y_{10z}$ modes. Given the richness of the observed pulsational behavior,
this star represents a unique laboratory through which we can investigate the
effects of a companion's gravitational field on the pulsations of
a star. We have also simultaneously
fit the SED and the binary lightcurve for the system parameters 
using an MCMC algorithm. This analysis showed that this system consists of a slightly
evolved primary $\delta$\,Sct star, with a secondary K-type star that is still on the
zero-age main sequence. 

We find that our best-fit inclination angle of $73.3^\circ\pm0.6^\circ$
is well outside the expected regime for the observed
doublets to be $Y_{\rm 11x}$ dipoles. This provides a useful
way to rapidly confirm future tri-axial pulsators; we also
provide a toy model framework in which this system and 
others can be interpreted in terms of tidal tilting. 
A targeted search for similar
pulsators is underway, with a specific focus on ellipsoidal
binaries \citep{green_ellipsoidal} and eclipsing binaries
that lie in the $\delta$\,Sct instability strip. Such a search
does not solely rely on the presence of eclipses, allowing
us to test our predictions even in the absence of eclipses, as
well as for a large range of orbital inclination angles. As the TESS
mission releases more light curves at 200\,s cadence, we will be able to
identify tri-axial pulsation in many more classes of stars.

\section*{Acknowledgements}

RJ would like to thank Kevin Burdge for discussions about light curve
fitting using {\tt Lcurve} \citep{lcurve}, which we (unfortunately!)
did not end up using as part of this work. RJ would also like to thank
Michael Fausnaugh for information about generating barycentric corrections 
for TESS light curves. 

V. B. K., S. R., and B. P. acknowledge financial support of the NASA 
Citizen Science Seed Funding Program, grant number 22-CSSFP22-0004.
GH thanks the Polish National Center for
Science (NCN) for supporting this study through grant
2021/43/B/ST9/02972.

This paper includes data collected by the TESS mission. Funding for the 
TESS mission is provided by the NASA Science Mission Directorate.
The QLP data used in this work was obtained from MAST
(\dataset[10.17909/t9-r086-e880]{https://dx.doi.org/10.17909/t9-r086-e880}),
hosted by the Space Telescope Science Institute (STScI).
STScI is operated by the Association of Universities for 
Research in Astronomy, Inc., under NASA contract NAS 5–26555.

This work also presents results from the European Space Agency (ESA) 
space mission Gaia. Gaia data are being processed by the Gaia Data 
Processing and Analysis Consortium (DPAC). Funding for the DPAC is 
provided by national institutions, in particular the institutions 
participating in the Gaia MultiLateral Agreement (MLA). The Gaia 
mission website is \url{https://www.cosmos.esa.int/gaia}. The Gaia archive
website is \url{https://archives.esac.esa.int/gaia}. This research has 
also made use of the VizieR catalogue access tool, CDS, 
Strasbourg, France. 

\facilities{TESS, Gaia}

\software{\texttt{astropy} \citep{astropy_2013,astropy_2018,astropy_2022},
         \texttt{lightkurve} \citep{lightkurve},
         \texttt{matplotlib} \citep{matplotlib},
         \texttt{numpy} \citep{numpy},
         \texttt{scipy} \citep{scipy},
         \texttt{synphot} \citep{synphot},
         \texttt{TICA} \citep{tica}
}

\bibliography{triaxial}{}
\bibliographystyle{aasjournal}

\appendix
\section{Numerical Integration for Limb-Darkening}
\label{app:ld}

As part of our light curve fitting code, we implemented a one-dimensional
numerical integral to account for limb darkening. For each
step in the light curve, we calculated the distance between the two stars'
centers $s$, and the distance from the primary star's center to the limb of the
secondary ($s - r_2$). We then split the range from $[s-r_2, r_1]$ into 75 steps,
which corresponded to 75 circles concentric with the primary, with radii
lying within that range; we denote these radii by $r_{\rm conc}$. 
For each of the circles, we calculated the distance from the center of 
the primary star to the chord between the circle's two points of intersection with 
the secondary's limb. This value is given by the expression
\begin{equation}
    d = \frac{s^2 - r_2^2 + r_{\rm conc, x}^2}{2s},
\end{equation}
where $x$ corresponds to a given concentric circle.

This value enabled us to calculate the angle $\theta$ subtended by the chord 
between the two points of intersection, as $\theta/2 = \cos^{-1}(d/r_{\rm conc, x})$. 
The expression for the area of the portion of an infinitesimally thin circular ring that subtends an
angle $\theta$ is given by $r\theta\,dr$. We then integrated this against the
limb-darkened flux $L(r)$, from $[s-r_2, r_1]$. The formal integral for the
limb-darkened flux (of the primary) within the overlapping area is given by

\begin{equation}
    \int_{s-r_2}^{r_1} 2r\,\cos^{-1}\left(\frac{d}{r}\right)\,L\left(\frac{r}{r_1}\right)\,dr
\end{equation}
The radius $r$ was normalized to $r_1$ in the argument of the LD function (given below), as
the limb darkening formula typically assumes a circle with radius scaled to 1.

\begin{equation*}
    L(r) = 1 - c_0(1-\sqrt{1-r^2}) - c_1(1-\sqrt{1-r^2})^2
\end{equation*}

This was then normalized with the total limb-darkened flux across the surface
of the star, and then subtracted from 1, to determine the fraction of flux from the
primary that was visible. To calculate the total limb-darkened flux, we numerically
integrated the following expression, with $\Delta r = 0.005$:
\begin{equation}
    \int_0^{r_1} 2\pi r\,dr\,L\left(\frac{r}{r_1}\right)\,dr
\end{equation}

For the secondary, whose flux was 1\% that of the primary,
we instead calculated the limb darkening at $(r_1 + r_2 - s)/2$ and
applied that across the entire overlapping area. We note that this approximation scheme
breaks down when the center of the secondary is occulted by the primary star; however, 
given that this scenario only occurs for a small number of phase bins in our system, and 
we do not observe any significant residuals in Figure \ref{fig:lc-mod-resid} as a result
of this approximation scheme, we opt to ignore this higher-order correction. Obscuration of the stellar 
center would correspond to cases VII and VIII in the scheme of \citet{mandel_agol}. 
\end{document}